\documentclass[%
 reprint,
 amsmath,amssymb,
 aip, 
 apl, 
showkeys,
]{revtex4-1}

\usepackage{graphicx}
\usepackage{dcolumn}
\usepackage{bm}
\usepackage{hyperref}

\usepackage{amsthm}

\newcommand{\comment}[1]{}
\newcommand{\Diff}{\mathit{Diff}}

\newtheorem{proposition}{Proposition}

\newtheorem{definition}{Definition}
\newtheorem{control-problem}{Control Problem}

\comment{
\title
{Variance Bounds on Estimation of the Canonical Volume Form via Diffeomorphic Mapping}

\author{Daniel J. Tward$^{(1)}$}
\address[1]{Center for Imaging Science, Johns Hopkins University, Baltimore, MD, 21218}

\author{Partha P Mitra$^{(2)}$}
\address[2]{Cold Spring Harbor Laboratory, Cold Spring Harbor, NY, 11724}

\author{Michael I. Miller$^{(3)}$}
\address[3]{Department of Biomedical Engineering, Johns Hopkins University, Baltimore, MD, 21218}

\email{dtward@cis.jhu.edu}
\date{\today}
}

\begin{document}
\preprint{AIP/123-QED}

\title{
Estimating Diffeomorphic Mappings between Templates and Noisy Data: Variance Bounds on the Estimated Canonical Volume Form
}

\comment{
\author{Ann Author}
 \altaffiliation[Also at ]{Physics Department, XYZ University.}
\author{Second Author}%
 \email{Second.Author@institution.edu}
\affiliation{%
 Authors' institution and/or address\\
 This line break forced with \textbackslash\textbackslash
}%

\collaboration{MUSO Collaboration}

\author{Charlie Author}
 \homepage{http://www.Second.institution.edu/~Charlie.Author}
\affiliation{
 Second institution and/or address\\
 This line break forced
}%
\affiliation{
 Third institution, the second for Charlie Author
}%
\author{Delta Author}
\affiliation{%
 Authors' institution and/or address\\
 This line break forced with \textbackslash\textbackslash
}%

\collaboration{CLEO Collaboration}
}

\author{Daniel J. Tward}
\email{dtward@cis.jhu.edu}
\affiliation{Center for Imaging Science, Johns Hopkins University, Baltimore, MD, 21218}
\author{Partha P Mitra}
\affiliation{Cold Spring Harbor Laboratory, Cold Spring Harbor, NY, 11724}
\author{Michael I. Miller}
\affiliation{Department of Biomedical Engineering, Johns Hopkins University, Baltimore, MD, 21218}

\date{\today}

\comment{
\begin{abstract}
An article usually includes an abstract, a concise summary of the work
covered at length in the main body of the article. 
\begin{description}
\item[Usage]
Secondary publications and information retrieval purposes.
\item[PACS numbers]
May be entered using the \verb+\pacs{#1}+ command.
\item[Structure]
You may use the \texttt{description} environment to structure your abstract;
use the optional argument of the \verb+\item+ command to give the category of each item. 
\end{description}
\end{abstract}

\pacs{Valid PACS appear here}
\maketitle


}

\begin{abstract}
Anatomy is undergoing a renaissance driven by availability of large digital data sets generated by light microscopy. A central computational task is to map individual data volumes to standardized templates. This is accomplished by regularized estimation of a diffeomorphic transformation between the coordinate systems of the individual data and the template, building the transformation incrementally by integrating a smooth flow field. The canonical volume form of this transformation is used to quantify local growth, atrophy, or cell density. While multiple implementations exist for this estimation, less attention has been paid to the variance of the estimated diffeomorphism for noisy data. Notably, there is an infinite dimensional un-observable space defined by those diffeomorphisms which leave the template invariant. These form the stabilizer subgroup of the diffeomorphic group acting on the template. The corresponding flat directions in the energy landscape are expected to lead to increased estimation variance.  Here we show that a least-action principle used to generate geodesics in the space of diffeomorphisms connecting the subject brain to the template removes the stabilizer. This provides reduced-variance estimates of the volume form. Using simulations we demonstrate that the asymmetric large deformation diffeomorphic mapping methods (LDDMM), which explicitly incorporate the asymmetry between idealized template images and noisy empirical images, provide lower variance estimators than their symmetrized counterparts (cf. ANTs). We derive Cramer-Rao bounds for the variances in the limit of small deformations. Analytical results are shown for the Jacobian in terms of perturbations of the vector fields and divergence of the vector field.

\end{abstract}
\keywords{computational anatomy, morphometry, cell density, Hamiltonian dynamics}

\maketitle


\section{Introduction}
Computational Anatomy (CA) is a growing discipline. Starting with initial work \cite{GrenanderMiller1998,GrenanderMiller2007} directed towards the
the study of transformations between anatomical coordinate systems suitable for volumetric images of the sub-compartments of the human brain acquired largely using MRI, contemporary applications have been extended to much larger data volumes acquired using light microscopy.
Infinite dimensional diffeomorphisms constitute the central transformation group for studying shape and form.
\cite{Younes2010,Trouve2011}
The diffeomorphism model underlying this analysis assumes that the space of measured MRI and optical imagery can be generated from exemplars or templates via diffeomorphic changes of coordinates.
At the mesoscopic scale the variation of the diffeomorphic change in coordinates from one brain to another can represent transverse individual variation, including pathological conditions, or longitudinal developmental variation.

Of particular interest in mesoscale neuroanatomy are quantities such as the spatial densities of cellular somata or neuronal processes.
Mapping estimates of these quantities to a template or reference space requires estimation of the change in the local scale as captured by the determinant of the metric tensor (the canonical volume form determined by the Jacobian determinant of the mapping). These applications point to the importance of uncertainty estimation for the diffeomorphic transformations involved. This is the subject area of the current paper.

Dense mapping between coordinate systems began with
the low-dimensional matrix Lie groups forming the basis of Kendall's shape theory.\cite{Kendall1989}
Their infinite dimensional analogue, the diffeomorphisms between coordinate systems, were first introduced by Christensen et al.\ \cite{Christensen1996}
and have occupied a central role in CA.
\cite{Trouve1995a,GrenanderMiller1998,Dupuis1998,Miller-Younes-2001,Toga2001,Miller2002,Thompson2002,Miller2004-growth,beg2004computational,joshi2004unbiased,Ashburner2007,GrenanderMiller2007,Durrleman2008b,Ashburner2009,younes2009evolutions,Vercauteren2009DiffeomorphicDE,Younes2010,Pennec2011,adams2013computational} 
The small deformation methods have been associated with Bookstein's landmark matching \cite{Bookstein:1989:PWT:66131.66134,bookstein1991thin,Bookstein:1996} and subsequently for image matching.\cite{Bajcsy-Lieberson-Reivich-1983,BajcsyKovacic1989,Gee93elastic,AmitGrenander1989} 
Large deformations were studied as simply topology preserving transformations without a metric structure.\cite{Christensen1996,Christensen97volumetrictransformation,Haller1997,trouve1998variational,karaccali2004estimating}
The symmetric approaches
for large deformations were variants of these methods.\cite{avants2004geodesic,beg2007symmetric,avants2008symmetric,vercauteren2008symmetric}
The large deformation diffeomorphic metric mapping algorithms (LDDMM) emerged corresponding to Lagrangian and Hamilton's principles applied to the flow fields incrementally generating the diffeomorphic transformations involved
\cite{JoshiMiller2000,CamionYounes2001,glaunes2004diffeomorphic,Beg2005,Vaillant05surfacematching,Miller2006,GlaunesQiu2008,ceritoglu2010large,Risser2011,Sommer2011,twardceritoglu2011,DuYounesQiu2012,VadakkumpadanTrayanova2012,Tang-Miller-Mori-2012,du2012diffeomorphic,du2014diffeomorphic,risser2013piecewise,Tward2013,khan2013multistructure,Tward-Miller-Trouve-Younes-2016}
and provide a metric between images and diffeomorphisms. 
The survey article by \citet{sotiras2013deformable} places these works in the greater context of deformable registration.
%

Uncertainty of nonrigid registration algorithms has been investigated for several applications such as spatially adaptive smoothing of population level data or uncertainty visualization for surgical planning.\cite{simpson2006using,kybic2008fast,risholm2010summarizing,kybic2010bootstrap,simpson2011longitudinal,simpson2012probabilistic} These approaches to uncertainty have typically used resampling techniques and linear elastic or spline based deformation models as opposed to diffeomorphisms.  Theoretical bounds on uncertainty have been investigated for the case of translation-only registration.\cite{ketcha2017effects}

In this paper we study problems associated with the estimation of the (local) canonical volume form (ie ${\sqrt{det(g)}}$ where $g$ is the metric tensor of the diffeomorphism) via methods based on large deformation diffeomorphic metric mapping (LDDMM) \cite{Beg2005,Vialard2012}.
As an example of an application where knowledge of this local scale change is important, consider mapping estimated cell densities from an individual brain to a template. Investigators may wish to distinguish differential changes in cell density of different types, from an overall change in scale given by the diffeomorphism relating the individual brain to the template. This requires an estimation of the volume form.  In human neuroanatomy, this local change in scale is commonly used to quantify patterns of tissue atrophy in neurodegenerative disease  \cite{ashburner2000voxel}.

In the notation followed in the paper, the coordinate transformation $\phi$ has  Jacobian $\left[D \phi\right]_{ij}  = \left( \frac{ \partial \phi_i }{\partial x_j} \right) $ and first fundamental form (metric tensor) given by 
\begin{equation*}
g(\phi) =  (D \phi )^T (D \phi ) \ .
\end{equation*}
The canonical volume form is given by the square-root of the determinant $\sqrt { | \det g |}$.

We show that the geodesic equations associated with the LDDMM method provide a crucial reduction, removing the nuisance dimensions of the  underlying symmetries associated with the stabilizer of the template, i.e. the subgroup of the diffeomorphic transformations that leave the deforming template unchanged.
 This template-centered reduction of the stabilizer gives rise to the asymmetric mapping properties central to LDDMM, and provides robustness when the imaging targets suffer from an incomplete and or noisy measurement process. We examine the mean square error of estimation of the
canonical volume form.
We demonstrate that the asymmetry of LDDMM coupled to geodesic reduction of indeterminate dimensions of the flow leads to favorable performance of this estimate in the presence of noise and variability when compared to symmetric methods for image matching originally proposed by
\citet{Christensen-Johnson-2001}, and \citet{Avants-grossman-gee}.  
We also calculate the Cramer-Rao bound for the variance of the volume form in the case of small deformations.

Explicitly studying the behavior of the canonical volume form under uncertainty, as contrasted to registration accuracy which has been addressed by the community in some situations, is essential for drawing meaningful conclusions about cell or process density and brain morphometry \cite{ashburner2000voxel}.  The current work provides a theoretical basis for the observations presented and for related observations by other authors. It clarifies important differences between symmetric and asymmetric methods as they relate to the uncertainty of the volume form estimates.

The paper is organized as follows:
\begin{itemize}
\item 
We first review the theoretical methods including two optimal control problems that are stated for retrieving the diffeomorphism and the fundamental forms describing changes of coordinate systems between images. 
\item
The necessary condition for the solution to the variational problem is stated in terms of the Euler-Lagrange equation on the conjugate momentum. It is shown that this equation incorporates a Hamiltonian reduction of the infinite-dimensional symmetry group corresponding to the stabilizer of the template. 
\item 
We then derive a new analytical result for the Cramer-Rao bound on the variance of the fundamental form in the presence of measurement noise. 
\item
Following this we show results from large deformation simulations showing the decrease in variance for the asymmetric LDDMM which privileges the template coordinate system as ground truth and removes the stabilizer with respect to the template. 
\item
These results are compared to symmetric methods derived by Christensen and Johnson\cite{Christensen-Johnson-2001} and Avants and Gee et al.\cite{Avants-grossman-gee} 
It is shown that the Hamiltonian-reduced LDDMM and symmetric methods behave similarly at low-noise but LDDMM outperforms with increasing noise. 
\end{itemize}

The significant new results in this paper thus address uncertainty estimation of the canonical volume form for estimated diffeomorphic transformations for both large deformation LDDMM and symmetric methods as well as small deformation methods.

\section{Methods}
Ethics approval is not required for this work.
\subsection{Theoretical Methods}
\subsubsection{Geodesic Flows of Diffeomorphisms for Dense Transformation of Coordinate Systems}
The diffeomorphism model in Computational Anatomy posits that the diffeomorphism group acts on templates $I_{\text{temp}}(x), x \in X \subset  {\mathbb R}^3$ via group action to generate the space of observed anatomical data from individual subjects $\mathcal I$,
\begin{equation}
\mathcal I = \{ I = I_{\text{temp}} \circ \phi_t^{-1} \ , \phi_t \in \Diff \} \ ,
\end{equation}
with diffeomorphisms $\phi_t:X  \to X, t \in [0,1], \phi_t \in \Diff$ generated via flows:
\begin{equation}
\dot \phi_t = v_t \circ \phi_t \ , \ \phi_0 = {\rm id} .
\label{ODE-equation}
\end{equation}
where $\textrm{id}$ is the identity transformation.  This is also termed the random orbit model: in group theoretic terms, $\mathcal I$ is the orbit of $I_{\text{temp}}$ under $\Diff$.

The Eulerian vector fields $v_t: X \to  {\mathbb R}^3$
are constrained to be spatially smooth, supporting at least 1-continuous derivative in space, ensuring that the flows are well defined and with smooth inverse.\cite{Dupuis1998}
They are modeled to be smooth with a finite
norm $(V,\|\cdot \|_V )$  defined by a differential operator $A: V \rightarrow V^*$,
$
 \| v\|_V^2 = \int_X Av \cdot v dx $.
It is conventional to use powers of the Laplacian for the differential operator $A$ with a sufficient number of generalized derivatives such that the flow fields are guaranteed to have at least 1 continuous derivative in space.\cite{Dupuis1998}
The kernel $K$ of the associated reproducing kernel Hilbert space is at least 1-time continuously differentiable in the spatial variables, and is given by the Green's kernel of $A$.


Mapping individual brains to reference atlases such as the Allen mouse atlas \cite{jones2009allen} at the micron scale or the Mori human atlas \cite{mori2008stereotaxic} at millimeter scale is performed via bijective coordinate transformation between anatomical coordinate systems $\phi$.
The diffeomorphic change in coordinates is not directly observable and must be inferred from observed brain image data subjected to measurement noise and technical variations. The coordinate system transformations may be estimated by solving the diffeomorphic matching problems as a solution of an optimal control problem. Different optimal control problems are obtained for large and small deformations. 

To set up the optimal control problem, we define the transformation $\phi_t$ as a time dependent state, $t \rightarrow \phi_t \in \Diff$. The velocity field
$t \rightarrow v_t$ is taken to be the control variable. The state satisfies the dynamical equations $\dot \phi_t = v_t \circ \phi_t $, with initial condition $\phi_0=\textrm{id}$. 
The goal of the optimal control problem is to drive the state from an initial condition of identity (corresponding to the template), to a state that matches template coordinates to target coordinates. This can be enforced by a cost function for the final state,
\begin{equation}
U(\phi_1) = \frac{1}{2\sigma^2} \int_X | J - I \circ \phi_1^{-1} |^2 dx \ .
\end{equation}
where $J$ is an observed target image. We assume that any differences in position or orientation have already been accounted for by an appropriate similarity transform before deforming the coordinate grid using the diffeomorphic procedure.  A scalar parameter $\sigma$ controls the importance of this data attachment term relative to the regularization term defined below.

Since there are an infinite number of possible flows we use the principle of least-action to minimize a running cost given by the integrated kinetic energy of the vector field  $v = \dot \phi \circ \phi^{-1}$ of the flow (LDDMM \cite{Miller2002,Beg2005}).
This regularization cost term can be interpreted as a kinetic energy Lagrangian:
\begin{align}
L(\phi_t,\dot \phi_t)&=\frac{1}{2}\int_X A (\dot \phi_t \circ \phi_t^{-1})\cdot \dot \phi_t \circ \phi_t^{-1} dx \notag \\
&=\frac{1}{2} \int_X Av_t \cdot v_t dx \ .
\label{Lagrangian-equation}
\end{align}

The cost function for the optimal control problem is obtained by combining the kinetic energy term (the regularization term) with an end-point cost. Thus, the optimal control problem involves Lagrangian mechanics of the infinite dimensional state defined by the coordinate transformation $\phi_t$. We study two versions of this problem, corresponding to large and small deformations (control problems 1 and 2).  

\begin{control-problem}[Large Deformation]
\label{sec:lddmm-formulation}

\[
\begin{array}{l} 
\displaystyle
\dot{\phi}_t= v_t \circ \phi_t \ , \phi_0 = \textrm{id}\\
\displaystyle
\min_\phi C(\phi) \doteq  \int_0^1 L(\phi_t,\dot \phi_t) dt   + U(\phi_1)
\end{array}
\]
\end{control-problem}
For small deformations
\cite{Bajcsy-Lieberson-Reivich-1983,BajcsyKovacic1989,AmitGrenander1989}
there is no time index, and $v$ represents a displacement field rather than a flow field, $\phi = id +v$. 
\begin{control-problem}[Small Deformation]

\[
\begin{array}{l} 
\displaystyle
\phi=\mathrm{id}+v \\
\displaystyle
\min_v C(\phi) \doteq \frac{1}{2} \int_X A v \cdot v dx  + U(\phi)
\end{array}
\]

\end{control-problem}

\begin{proposition}
The solution to the large deformation Control Problem 1 has classical conjugate momentum  $ p_t = Av_t \circ \phi_t |D \phi_t |$  \cite{Miller-Trouve-Younes-2014,Miller2015} which satisfies geodesic equations
\begin{equation}
\label{LDDMM-image-matching-problem2}
p_t = \alpha_t \nabla I , \ \ \ 
\alpha_t = \frac{1}{\sigma^2} (I-J\circ \phi_1) |D \phi_1| (D \phi_t^T)^{-1} \ .
\end{equation}

For small deformations, Control Problem 2,
$p=Av$ satisfying
\begin{equation}
\label{momentum-small-deformation}
p= \alpha \nabla I \ , \ \alpha=
\frac{1}{\sigma^2} (I-J\circ \phi )|D \phi |(D \phi^T)^{-1}  \ . 
\end{equation}
\end{proposition}
The conjugate momentum satisfying LDDMM 
equation \eqref{LDDMM-image-matching-problem2}
is perpendicular to the level lines of the template following the image gradient.
This normal condition is satisfied over the entire path of deformation.  This is observed most easily by considering the \emph{Eulerian momentum}, related to the conjugate momentum by a change of coordinates, $Av_t = \frac{1}{\sigma^2}|D\phi_{1t}^{-1}|(J(\phi_{1t}^{-1}) - I(\phi_t^{-1}))\nabla [I(\phi^{-1}_t)]$, where $\phi_{1t} = \phi_t(\phi_1^{-1})$.  Notice that $Av_t$ is parallel to $\nabla [I(\varphi_t^{-1})]$ at every point.
See Appendix \ref{Euler-Lagrange-appendix} for a proof.

The fact that the conjugate momentum is in the range space at every point of the gradient of the template is a Hamiltonian reduction that removes the symmetries of the template given by the stabilizer subgroup. The non-identifiable motions that are tangent to level lines are thus removed.
This kills off the nuisance dimensions, essentially suppressing components in the null-space of level lines of the template as it flows.\cite{Miller2006}

For small deformation matching the normal condition of the stabilizer corresponds to a single vector condition that $Av$ is in the span of $\nabla I$ the gradient of the image template.
Small deformations allow us to both calculate the Cramer-Rao bound (see below) as well as perform
a direct perturbation argument. 
Via a perturbation $J \rightarrow J +\delta  J $ we can calculate
the accuracy of the divergence of $v$ and how it determines the Jacobian determinant and the canonical volume measure related to the gradient of the template.

\begin{proposition}
For deformations close to the identity, a perturbation $J \mapsto J + \delta J$ results in a perturbation to the canonical volume form:
\begin{align}
\label{eq:small-perturbation}
\begin{cases}
\delta v &= [A + \frac{1}{\sigma^2} (\nabla I)  (\nabla I)^T]^{-1}\nabla I \delta J\\
|D\phi| &\mapsto |D\phi| + |D\phi|\text{div}[\delta v]
\end{cases}
\end{align}
where $\nabla I$ is a column vector, the outer-product giving a $3\times 3$ matrix.  
\end{proposition}
See Appendix \ref{Proof-of-small-deformation-momentum} and \ref{proof-of-small-perturbation} for small deformation proofs.  
Equation
\eqref{eq:small-perturbation} clearly shows the crucial matrix involving the gradient of the image and the prior represented via the differential operator $A$ for determining the 
perturbation on the canonical volume form through the divergence.
The Cramer-Rao bound reflects this theme.

\subsubsection{Cramer-Rao Bound (CRB) for Small Deformations}
\label{sec:crbound}
We now examine the variational estimator and the variance bound for small deformations in the finite dimensional setting of $n$-dimensional vector fields $ v^n(x) = \sum_{i=1}^n \theta^i \psi^i(x)$. Here $\psi^i$ is some suitable family of expansion functions. For computing the CRB, 
we take the observed data to be a conditionally Gaussian random field, conditioned on the mean $I \circ \phi^{n-1}$ with additive noise
\begin{equation}
J= I \circ (\phi^{n})^{-1} + noise  \ , \ \ \phi^n = \textrm{id} + v^n \ .
\end{equation}
The noise is taken to be zero-mean with non-white inverse covariance $Q$. 
The log-likelihood on the $n$-dimensional cylinder $\Theta^n = (\theta^1,\dots,\theta^n)$ is
\begin{align}
\ell_n (J;\Theta_n) 
&=
-
\frac{1}{2} \int_X\int_X (J(x)-I \circ \phi^{n -1 }(x)) \notag \\
&\qquad Q(x,y)(J(y)-I \circ \phi^{n -1 }(y)) dx dy \
\  .\label{log-likelihood-cylinder}
\end{align}
For
white-noise the inverse-covariance is $Q(x,y) = \frac{1}{\sigma^2 (x)} \delta(x-y)$ the $n$-dimensional log-likelihood \eqref{log-likelihood-cylinder}, for $\sigma^2(x)$ a variance at location $x$.
Adding the finite-dimensional prior term $- \frac{1}{2} \int_X Av^n(x) \cdot v^n(x) dx  $
gives a proper maximum a-posterior estimator (MAP) on $n$-dimensions:
\begin{align}
&\max_{\theta^1, \dots , \theta^n }
\log \pi^n (\Theta^n  | J ) \notag \\
&= - \frac{1}{2} \int_X Av^n \cdot v^n dx + \ell^n (J; \theta^1,\dots, \theta^n)
\ . 
\end{align}
The Fisher-information is the $n \times n$ matrix, $i,j=1,\dots,n$:
\begin{align*}
\mathcal I_{ij}(\Theta_n) &= E_{J|\Theta^n} \left\{  \frac{\partial \ell^n(J;\Theta^n)}{\partial \theta^i}  \frac{\partial \ell^n (J;\Theta^n)}{\partial \theta^j} \right \}
\ .
\end{align*}
Taking expectation over $\Theta^n$ gives the Bayesian version of the Fisher-information: 
$$\mathcal I^B = \int_{R^n} \mathcal I(J;\Theta^n) \pi(d \Theta_n)  \ . $$   
\begin{proposition}
Defining for all $(x,y)$,
$ Q^\phi (x, y ) = Q( \phi(x), \phi(y) |D \phi (x) | | D \phi (y) |$,
\begin{align}
&\mathcal I_{ij}(\Theta_n)
= \int_X \int_X   \psi^{iT}(x)[D\phi(x)]^{-T} \nabla I(x) \notag \\
&\qquad Q^\phi (x,y)  \nabla  I^T(y)[D\phi(y)]^{-1}   \psi^j(y)dxdy \ 
\label{CRB-equation}
\ .
\end{align}
Neglecting quadratic terms using small deformations, noting that the linear terms have an expected value of 0, and taking white noise variance $\sigma^2$  gives
\begin{align}
\mathcal I^B_{ij} &= \int_X \psi_i^T(x) \left(  A  +  \frac{1}{\sigma^2(x)}   \nabla I(x) \nabla  I^T(x)  \right) \psi_j(x) dx \notag \\
&\qquad + E[o(v^2)] \ ,
\label{Bayes-bound-equation}
\end{align}
giving the lower bound on the estimator:
\begin{align*}
\text{Cov}[\hat \Theta_n] \geq  [\mathcal I^B]^{-1}
\end{align*}
\end{proposition}
Noteworthy is
the CR bound \eqref{Bayes-bound-equation} is at the core the same form as direct perturbation
\eqref{eq:small-perturbation}.
We are particularly interested in the variance of $\text{div}(\hat v)$, because it directly relates to variance of the Jacobian.  This can be written as a linear functional of $v$, using a test function $w$ which is nonzero in a small neighborhood of the point $y$ (i.e. the sensitivity of an image voxel), namely $\int_X w(x)\text{div}v(x)dx $.
In this case the information inequality, which is concerned with estimating functions of the parameter $v$, becomes
\begin{align}
\label{eq:CRB}
\text{Var}[\text{div}[\hat v](y)] &\geq \sum_{i,j=1}^n \left( \int_X w(x) \text{div}\psi_i(x) dx\right) \notag \\
&\qquad [\mathcal I^B]^{-1}_{ij}\left( \int_X w(x) \text{div}\psi_{j}(x)dx\right)
\end{align}


\subsubsection{The Stabilizer: The normal condition for the nonlinear group action}
Unfortunately there is an infinite dimensional
nuisance parameter, termed the stabilizer, which is not uniquely determined in estimating the coordinates of the bijections between coordinate systems.
The geodesic motion kills the nuisance parameter, suppressing components in the instantaneous null-space of level lines of the template as it flows \cite{Miller2006};
this is Theorem 4 of \cite{Miller2015}.
Notice there is no momentum tangent to the level lines of the flow of the template image, $p_t = \alpha_t \nabla I $ in 
\eqref{LDDMM-image-matching-problem2}.
This is not true for other diffeomorphic mapping methods.
It results from the metric property of the geodesic equation
with it's associated conservation law. 

To understand flows which are normal to the level lines and in the span of the gradient of the template,
examine the stabilizer. 

\begin{definition}
Define the subgroup
$\mathcal S \subset \Diff $ as the stabilizer of template $I$ if for all $\phi \in  \mathcal S$,
$$
I \circ \phi = I \ . $$
\end{definition}

\begin{figure*}
\centering
\includegraphics[width=0.30\textwidth,clip,trim=0in 1in 0in 1in]{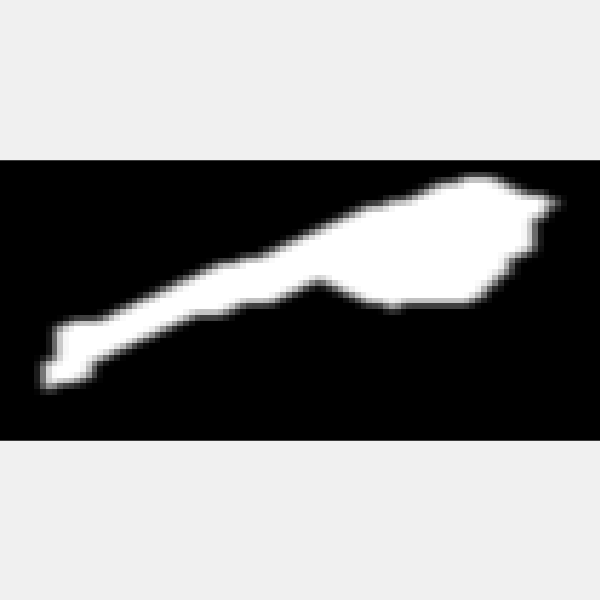}
\includegraphics[width=0.30\textwidth,clip,trim=0in 1in 0in 1in]{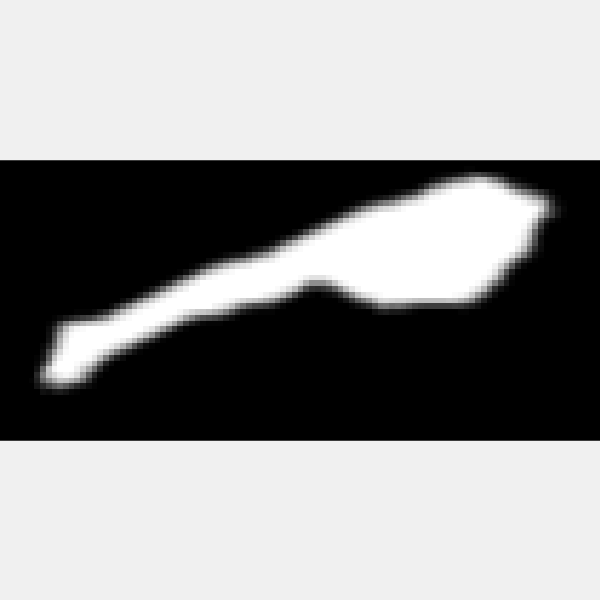}
\includegraphics[width=0.30\textwidth,clip,trim=0in 1in 0in 1in]{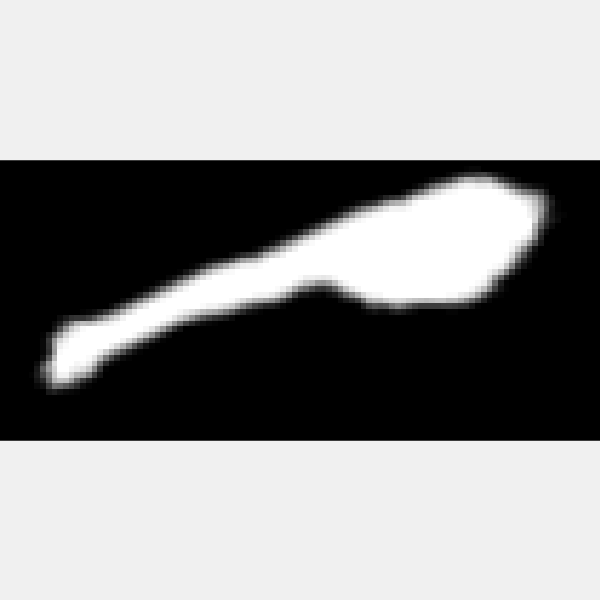}
\\
\includegraphics[width=0.30\textwidth,clip,trim=0in 1in 0in 1in]{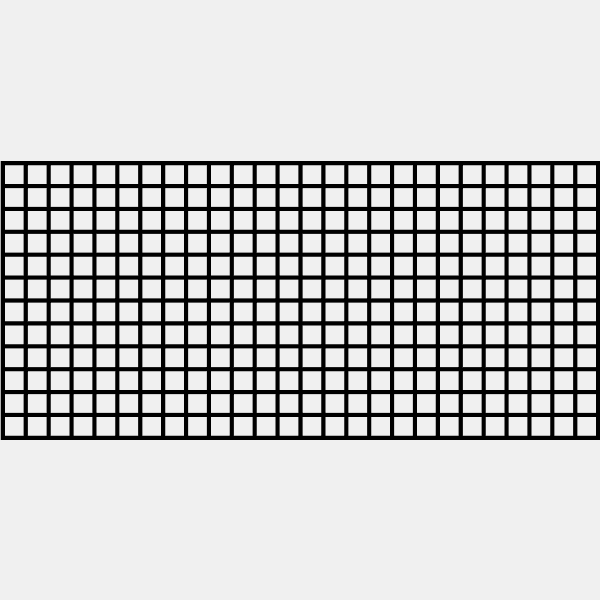}
\includegraphics[width=0.30\textwidth,clip,trim=0in 1in 0in 1in]{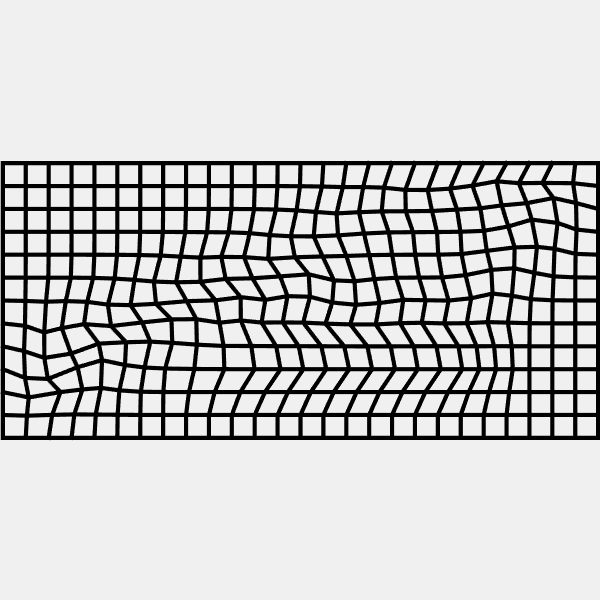}
\includegraphics[width=0.30\textwidth,clip,trim=0in 1in 0in 1in]{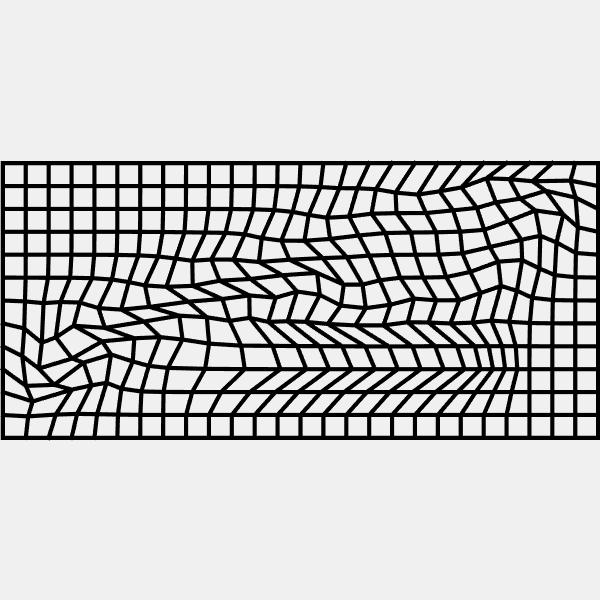}
\caption{\label{fig:stabilizer}Mappings using elements from the stabilizer, leaving the template 2D sections essentially unchanged (top row) but moving the coordinates (bottom row). 
Column 1 is identity $\phi=id$; columns 2,3 have a tangent component
$w \in V_{I }$ in the stabilizer. }
\end{figure*}

Figure \ref{fig:stabilizer} shows examples of mappings from the stabilizer of an image of a human hippocampus, to a close numerical approximation. The left column shows the identity mapping on the grid; the right two columns show two mappings from the stabilizer.

Vector fields that are normal to the level lines in the span $\nabla I $ do not generate flow through the stabilizer group.
The stabilizer is generated from flows $\dot \phi_t = w \circ \phi_t , \phi_0=id $ for which the vector fields are
tangent to level lines of the template:
\begin{align*}
V_I = \{ w \in V : \nabla I \cdot w = 0 \}  \ . 
\end{align*}
That being normal to the level lines is a necessary condition to null out the stabilizer, examine 
$\phi_\epsilon = id + \epsilon w $, then
\begin{align*}
 I \circ \phi_\epsilon = I  + \epsilon \nabla I \cdot  w + o(\epsilon) \ . 
\end{align*}
For $I \circ \phi_\epsilon = I$ it must be the case that $ \nabla I \cdot w = 0 $, implying $w$ must be normal
to the level lines.
This is the group action version of the pseudo-inverse condition for inverting a matrix with non-zero null-space.

\subsection{Numerical Experiments}
\subsubsection{Simulating Variance Bounds on Large Deformations}
The geodesic properties of LDDMM imply lower variance estimates of the first fundamental form and the fundamental volume measure.
To illustrate this we create simulated images by generating large deformation diffeomorphisms under the random orbit model,  deforming a template image under these known transformations, and applying Gaussian noise.  Our template corresponds to the binary segmentations of the anatomical section in Fig. \ref{fig:white-noise} (left column),  a sagittal section of an ex vivo image of the human medial temporal lobe, where hippocampal subfields and surrounding areas are visible.  
We generate random Gaussian vector fields $v_0(x) = v^n(x), x \in X$ as the initial condition of geodesic solutions for deformation.
Since geodesics satisfy a conservation law 
\begin{equation}
\label{conservation-law-conjugate-momentum}
\frac{d}{dt} D \phi_t^T p_t = 0
\end{equation}
proved in Appendix \ref{app:EL} equation \ref{conservation-conjugate-momentum}, we can synthesize the initial condition $p_0 = Av^n$ and generate a random spray of deformations $\phi_1$.
The Gaussian vector field is
generated from an $n$-dimensional expansion $\psi_i:  X \to \mathbb R^3,i=1,2,\dots$,
\begin{align*}
v^n(x) &= \sum_{i = 1}^n \theta^i \psi^i(x) \ ,
\end{align*}
coefficients $\theta^i \in \mathbb R, i = 1,\dots,n$ distributed as multivariate Gaussian, zero-mean $\bar \theta^i = 0$ and covariance $E[\theta^i \theta^j] = \Sigma^{ij}$.
We choose the covariance of vector fields $v^n $ to be well defined for large $n$,  sampled at pairs of points $x,y \in \mathbb R^3$, 
\begin{align*}
K^n(x,y) &\doteq E[v^n(x)v^{nT}(y)] = \sum_{i,j} \psi^i(x) E[\theta^i \theta^j] \psi^{jT}(x) \ .
\end{align*}
For large $n$ this tends to covariance specified by the Green's kernel of the operator $A$ originally stated by Beg
\cite{Beg2005},
 the inverse of $A = (1 -a^2\Delta)^4$ for $\Delta$ the Laplacian at spatial scale of $a = 0.25 \, \text{mm}$, or 2 pixels).  To achieve this, we choose an expansion $\psi^i$ that corresponds to a superposition of Green's kernels located on each voxel where the image gradient is nonzero (3432 expansion functions).
  
For the statistical characterization, 100 realizations of random deformation are generated running equation \eqref{conservation-law-conjugate-momentum} from $v^n$ as initial condition.  100 randomly generated images were constructed from these deformations applied to our template (Fig. \ref{fig:white-noise} left), embedded in additive Gaussian noise with standard deviation from 0 to 0.5, spatial correlation of 0 (white noise) or 1.5 pixels (obtained by convolving with a Gaussian of standard deviation 1.5 pixels).  Correlated noise is common in medical imaging systems such as radiography \cite{siewerdsen1997empirical}, CT \cite{tward2008cascaded}, or MRI \cite{gudbjartsson1995rician}.  Similarly, 100 randomly generated images were constructed using the identity transformation instead of a random transformation.

Applying diffeomorphic mapping for 100 images produced using the identity transformation ($\theta_i=0 \forall i$), and 100 images produced using these known transformations then gives variance estimates of the diffeomorphic mapping methods.  For contrast we compare LDDMM to the symmetrized version discussed in \cite{Christensen-Johnson-2001} ANTs which does not exploit the reduced representation as symmetry supports momentum on both the template and target:
\begin{align} 
&\max_{v: \phi^v = \int_0^1 v_t \circ \phi_t dt + id}
-\frac{1}{2} \int_0^1 \|v_t\|^2_V dx - \frac{1}{2\sigma^2}\|I\circ\phi^{-1} - J\|^2_{L_2} \notag \\
 &\qquad\qquad -\frac{1}{2} \int_0^1 \|v_t\|^2_V dx  - \frac{1}{2\sigma^2}\|I - J\circ \phi\|^2_{L_2} \ . 
\label{symmetric-algorithm}
\end{align}
Gradient descent is used to determine the optimal $v$, giving estimates of  $\hat \phi$ and the canonical form $|D \hat \phi |$. We use exactly the same parameters for both algorithms.

Parameters were fixed or varied experimentally as summarized in Table \ref{tab:experiment} in the Appendix \ref{experimental-parameter-table}.  In our computational implementation we use two constants in our objective function to be minimized:  $\frac{1}{\sigma^2_V}$ multiplying the regularization (kinetic energy) term, and $\frac{1}{\sigma^2_I}$ multiplying the data attachment term.  Since we are estimating the optimizer, but not the optimum, this is equivalent to choosing a single $\frac{1}{\sigma^2} = \frac{\sigma^2_V}{\sigma^2_I}$.

\subsubsection{Calculating the Cramer Rao Bound}
Finite dimensional cylinders were used to illustrate the Cramer Rao lower bound on the divergence of $\hat v_n = \sum_{i=1}^n \theta_i \psi_i$ from  \eqref{eq:CRB}.  Expansion functions are Green's kernels of $A$ located on image boundaries, where the image is downsampled by a factor of (from left to right) 8 (76 expansion functions), 4 (318 expansion functions), 
or 1 (not downsampled, 3432 expansion functions).  Explicitly inverting the ill-conditioned matrix $\mathcal I^B$ was avoided by solving the linear system implied by \eqref{eq:CRB}, which was performed using Matlab's \texttt{linsolve} for positive definite symmetric matrices.

\section{Results}
\subsection{Variance bounds}
\begin{figure*}
\centering
\newcommand{\statistic}{rmse}
\includegraphics[width=0.32\textwidth]{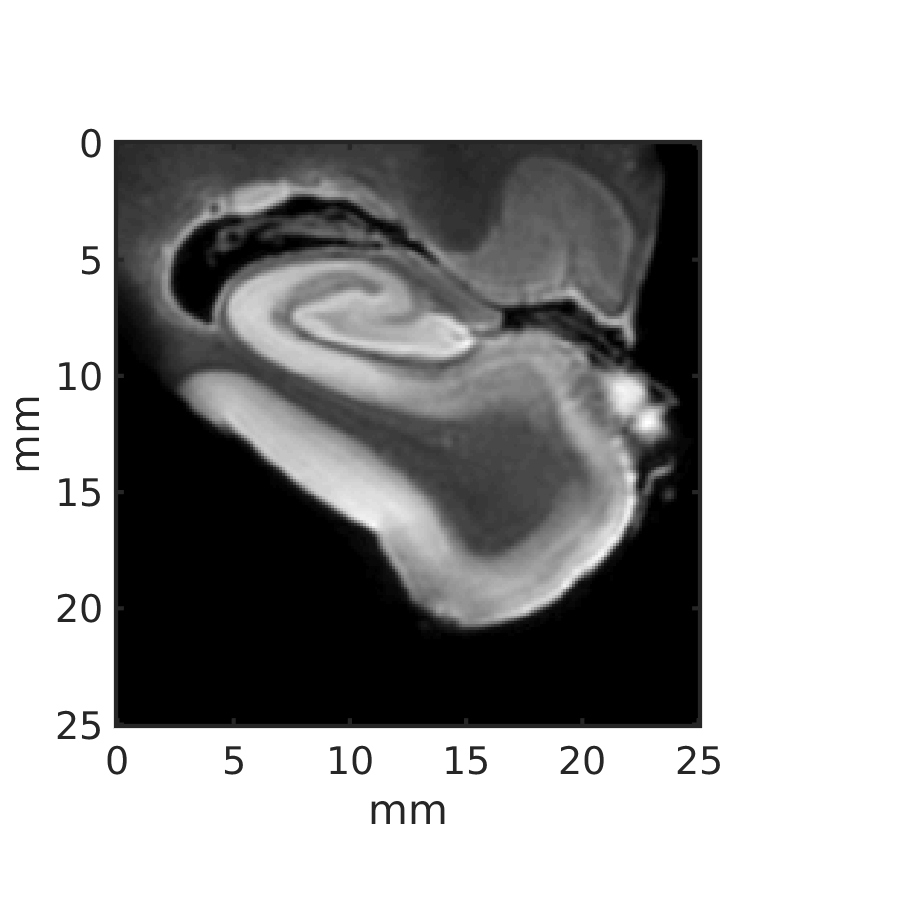}
\includegraphics[width=0.32\textwidth]{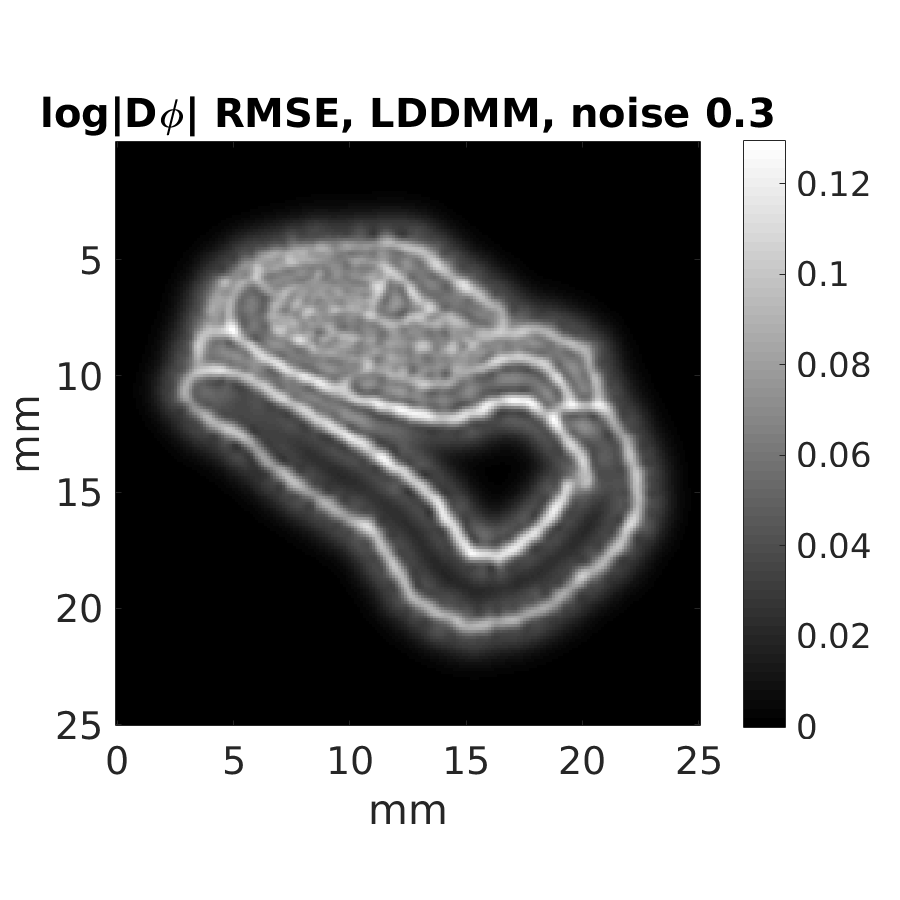}
\includegraphics[width=0.32\textwidth]{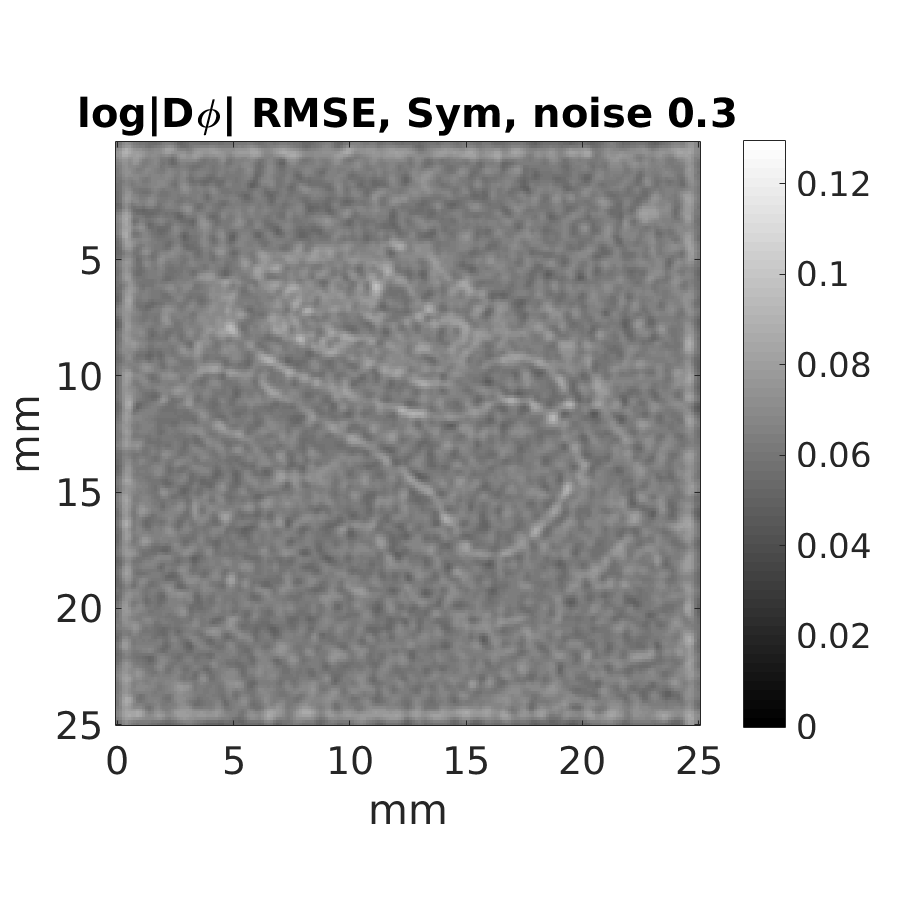}\\
\includegraphics[width=0.32\textwidth]{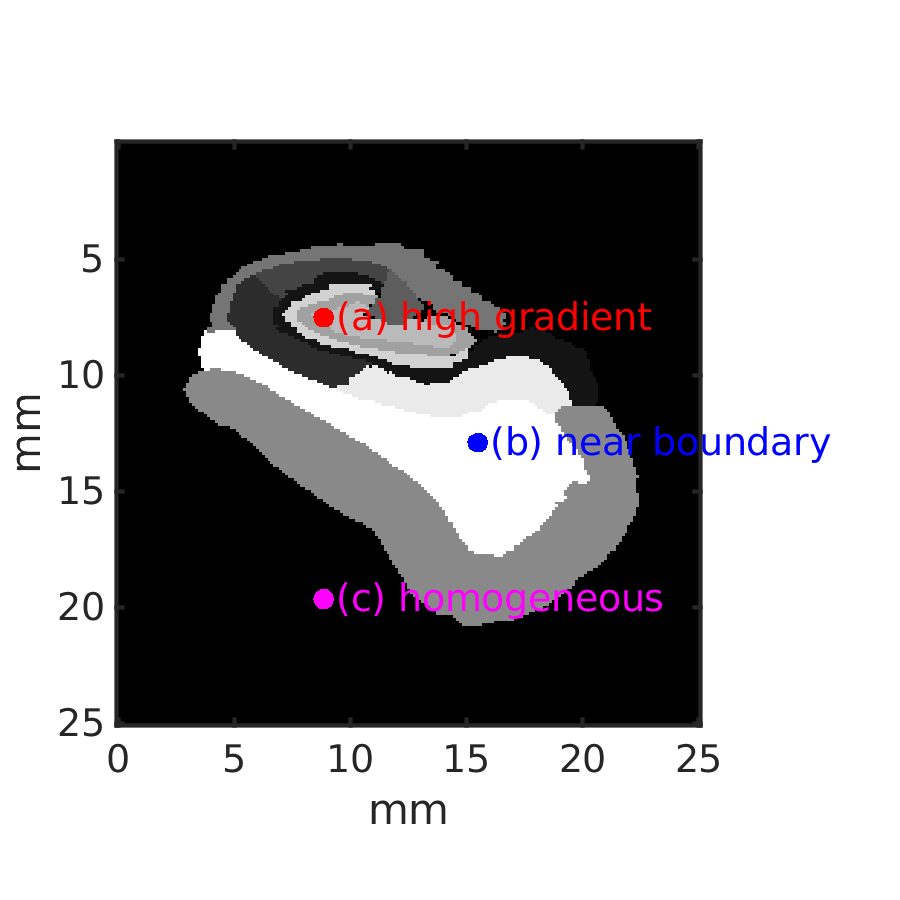}
\includegraphics[width=0.32\textwidth]{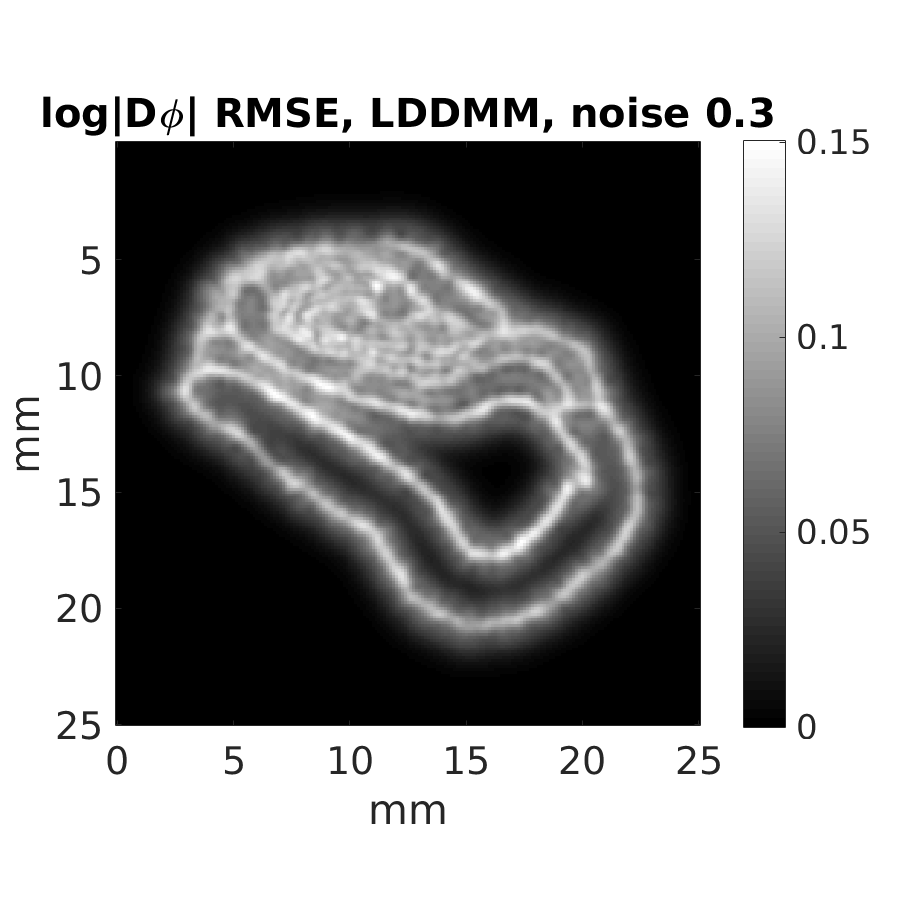}
\includegraphics[width=0.32\textwidth]{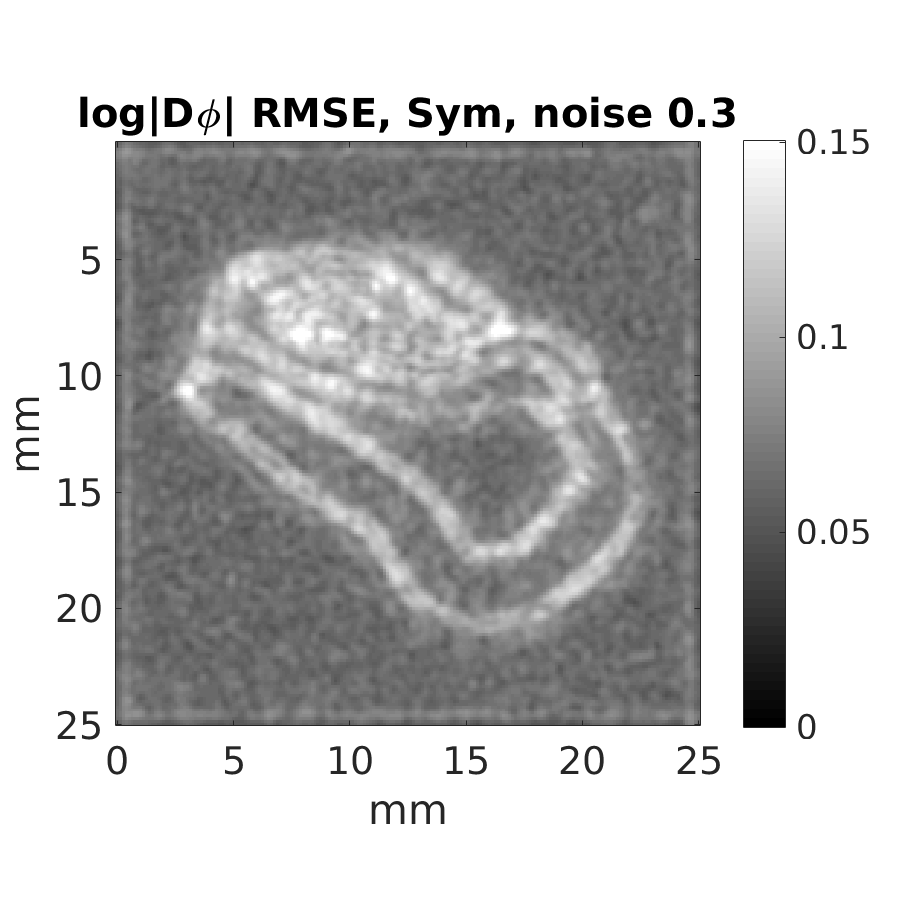}
\caption{
\label{fig:white-noise} 
Column 1 shows sections through MRI (top) of entorhinal cortex and subiculum and CA partition of the hippocampus (bottom) with three locations depicted for RMSE comparisons.
Column 2 shows RMSE's for canonical volume forms for the MTL section with
showing LDDMM method and
column 3 showing the symmetrized algorithm \eqref{symmetric-algorithm}.
Top row shows identity transformation;
bottom row shows the randomized transformation.
 }
\end{figure*}

Figure \ref{fig:white-noise} shows sections (column 1) through the high-field medial temporal lobe MRI phantom used for numerical experiments, with the segmentation into the entorhinal cortex and it substructures subiculum and Cornu Ammonis (CA) compartments.
Figure \ref{fig:white-noise} (columns 2,3) show results for simulating known transformations of the template with additive white noise.
Column 2 shows the results of solving LDDMM equation \eqref{LDDMM-image-matching-problem2}
to estimate each random deformation $\phi$. The top row shows identity transformation of coordinates; the bottom row shows random deformations $\phi$.
LDDMM in homogeneous regions has RMSE error which drops to nearly zero across the entire image, with more
error incurred at the contrast boundaries. Column 3 shows the symmetric case which has significantly higher RMSE across the image than LDDMM.  

\begin{figure*}
\centering
\includegraphics[width=0.324\textwidth]{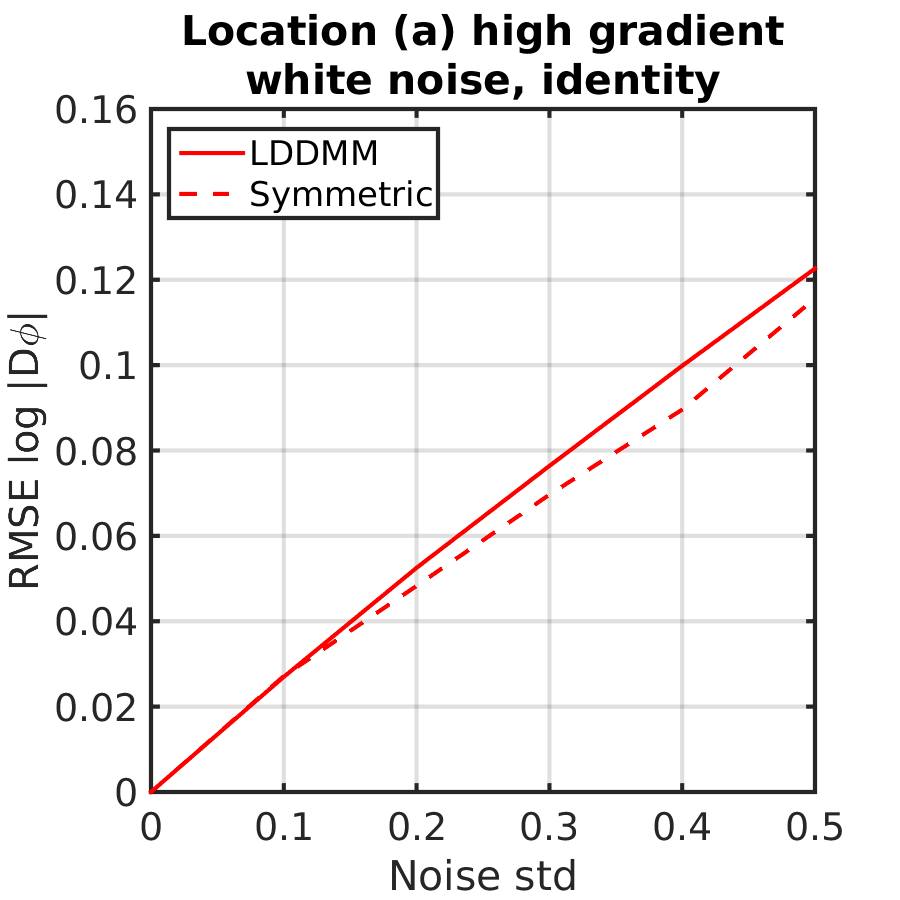}
\includegraphics[width=0.324\textwidth]{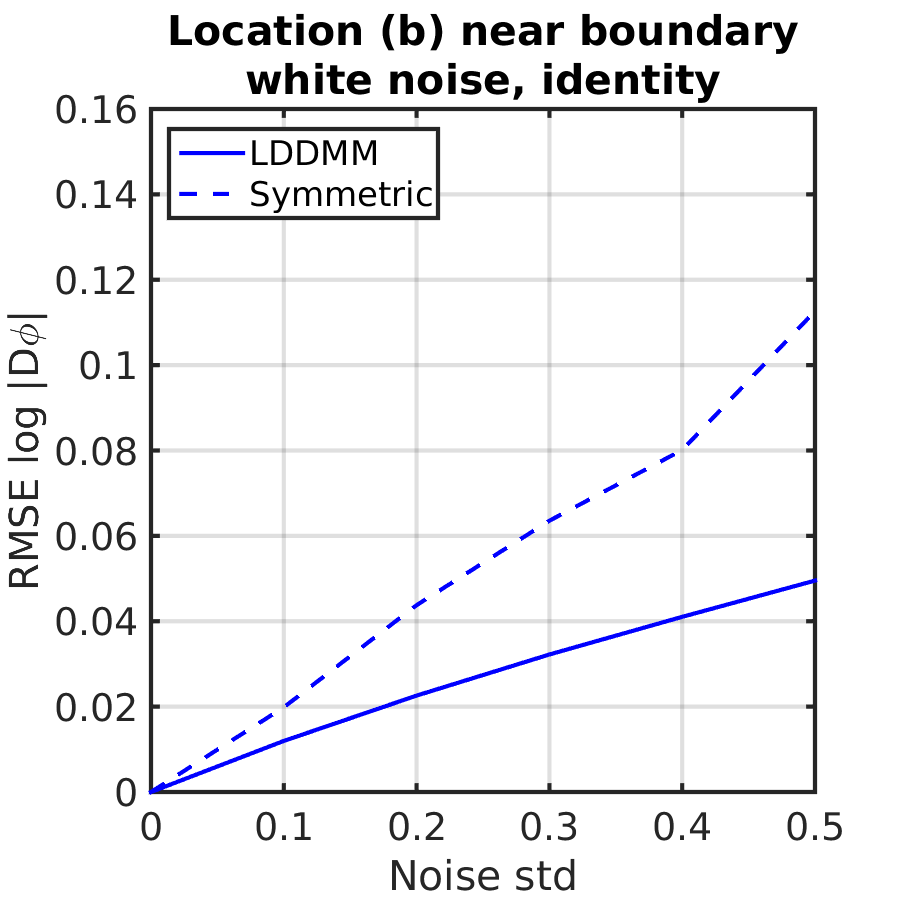}
\includegraphics[width=0.324\textwidth]{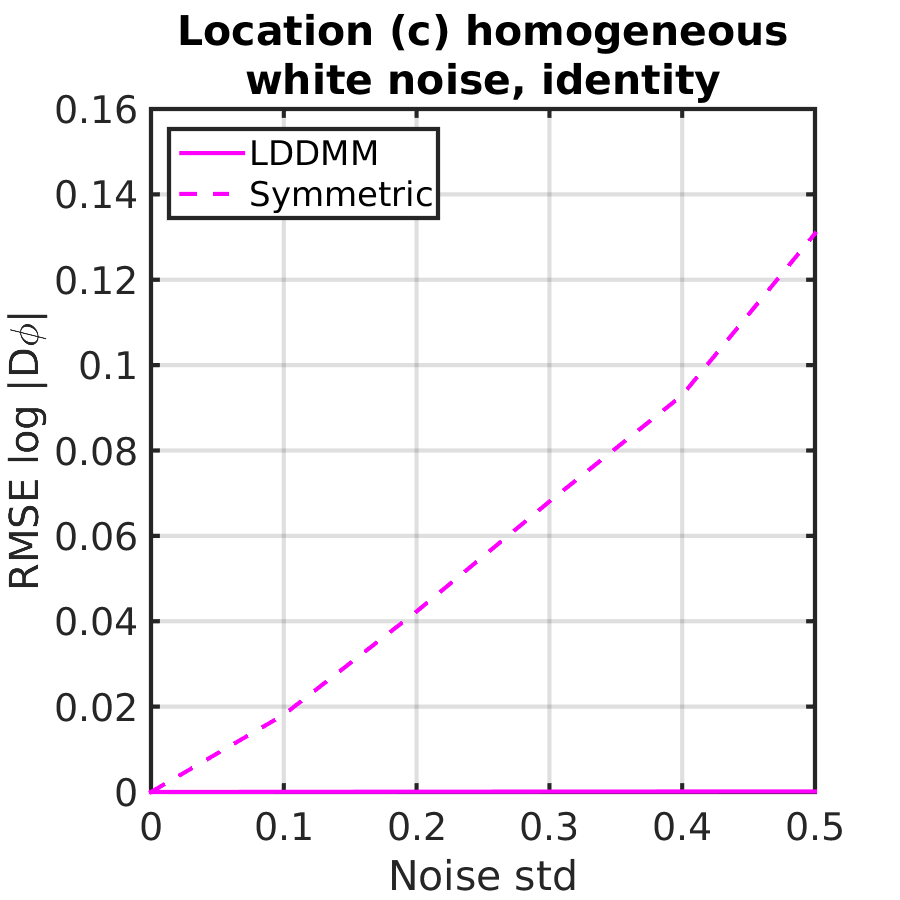}
\\
\includegraphics[width=0.324\textwidth]{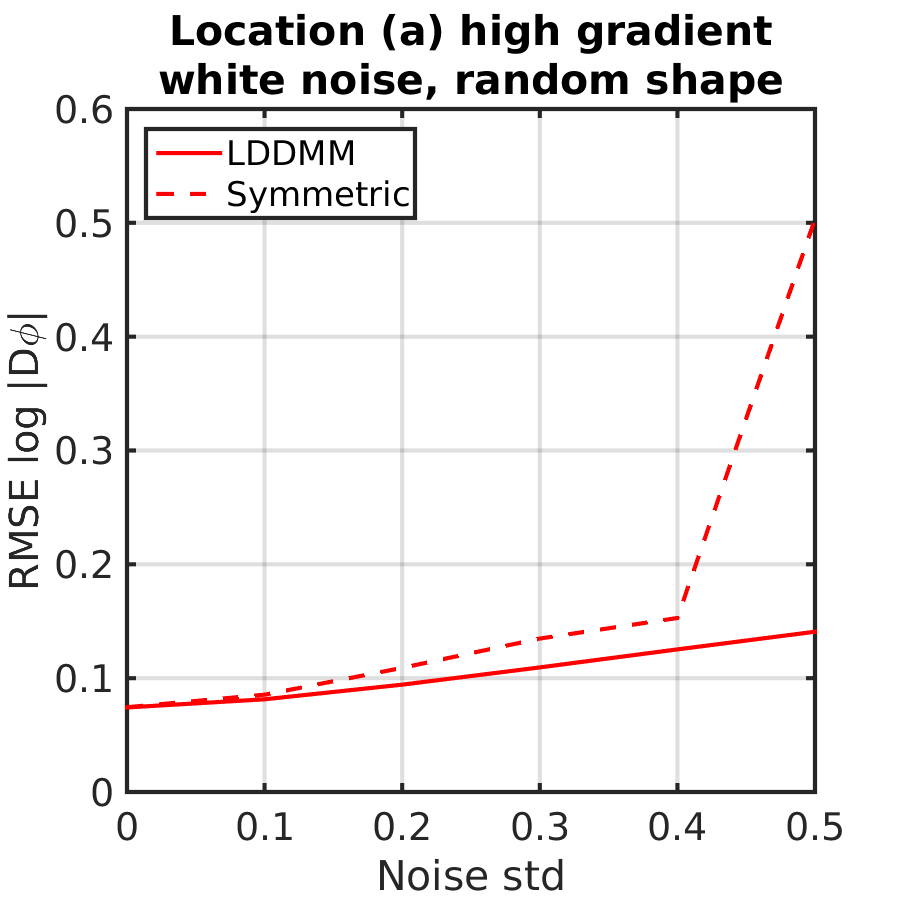}
\includegraphics[width=0.324\textwidth]{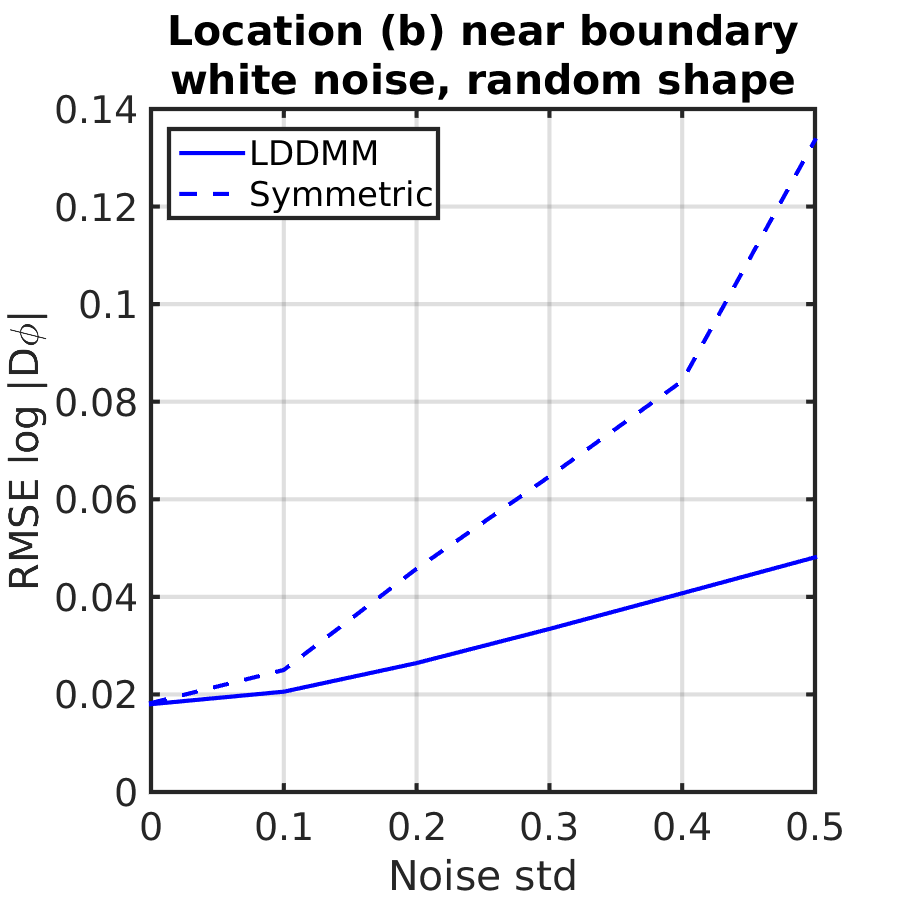}
\includegraphics[width=0.324\textwidth]{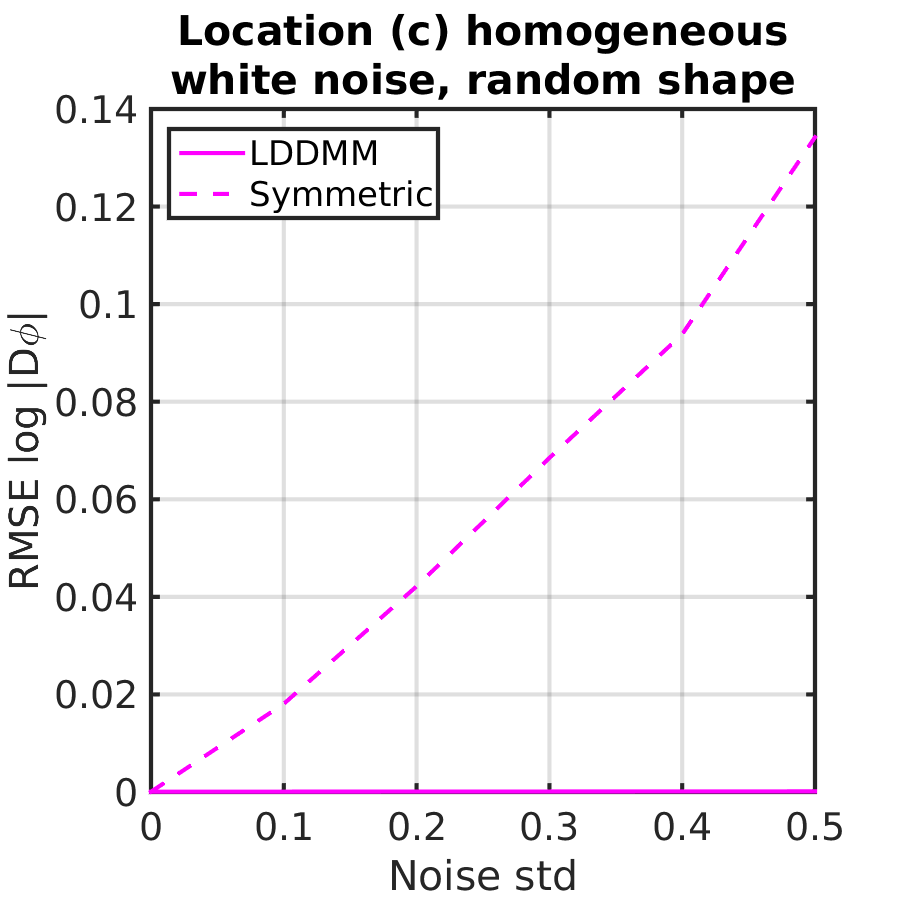}
\caption{\label{fig:curves-white-identity} \label{fig:curves-white-random} RMSE of log canonical volume form in white noise; 
LDDMM solid, symmetric dashed. Top row shows identity transformation; bottom row shows random transformation. Notice panel 1 shows very close performance of two methods; otherwise noteworthy differences. 
}
\end{figure*}

\begin{figure*}
\centering
\includegraphics[width=0.324\textwidth]{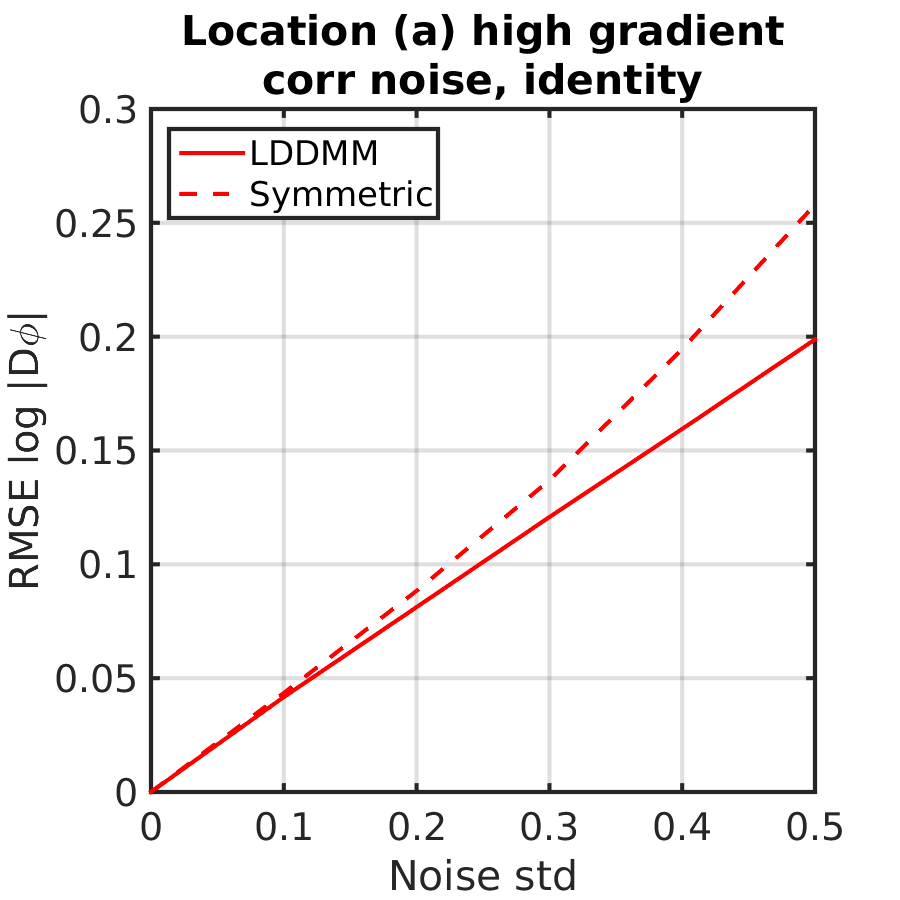}
\includegraphics[width=0.324\textwidth]{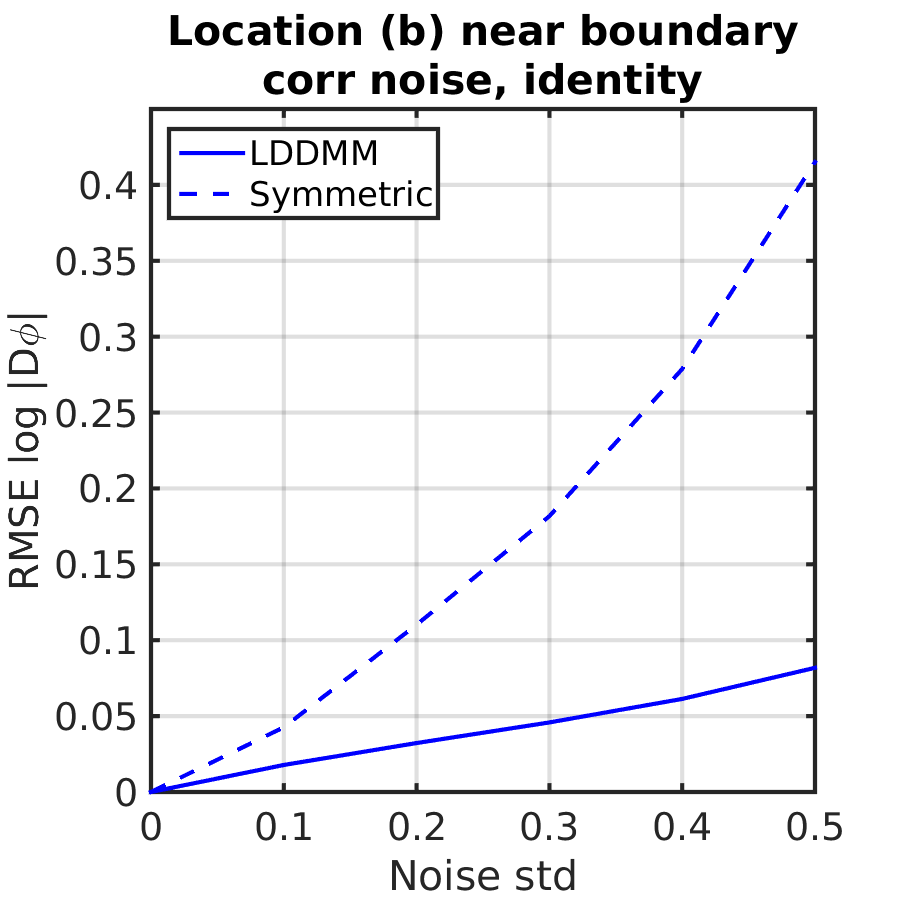}
\includegraphics[width=0.324\textwidth]{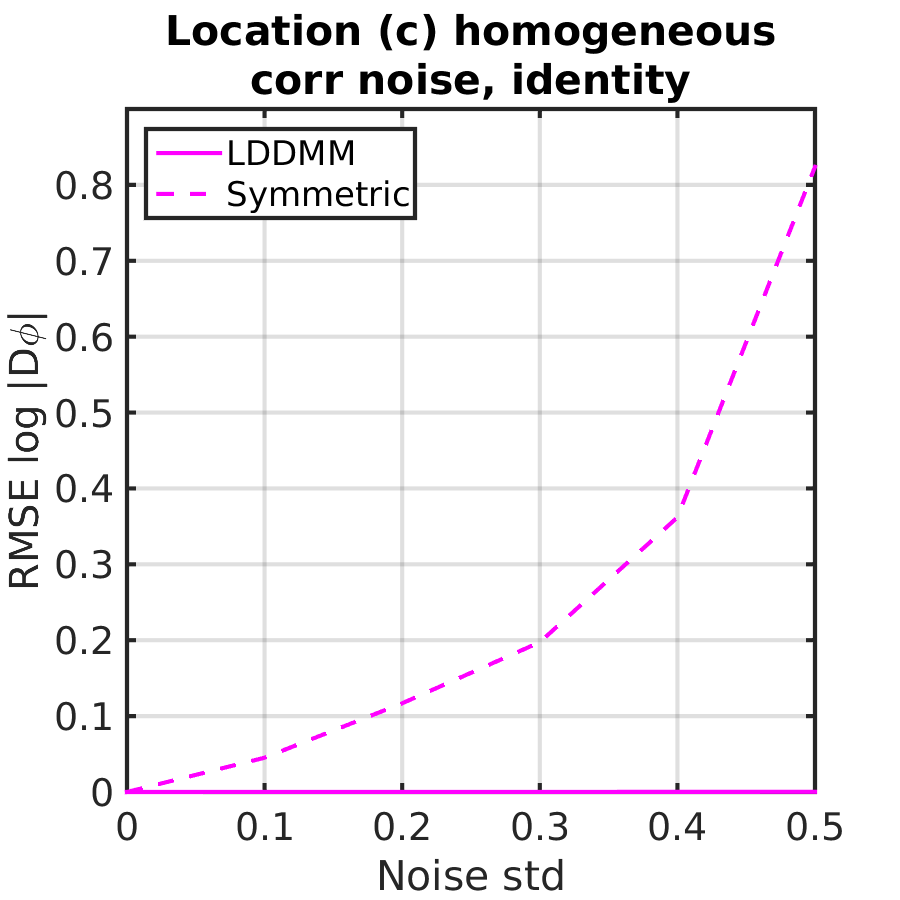}
\\
\includegraphics[width=0.324\textwidth]{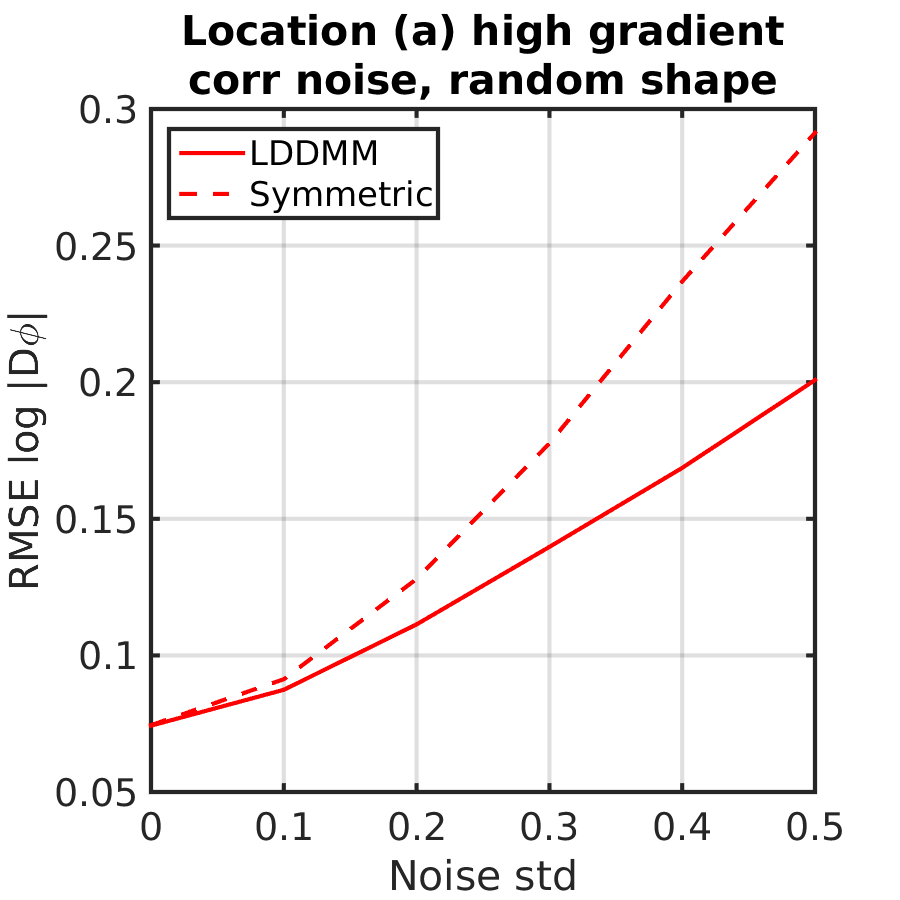}
\includegraphics[width=0.324\textwidth]{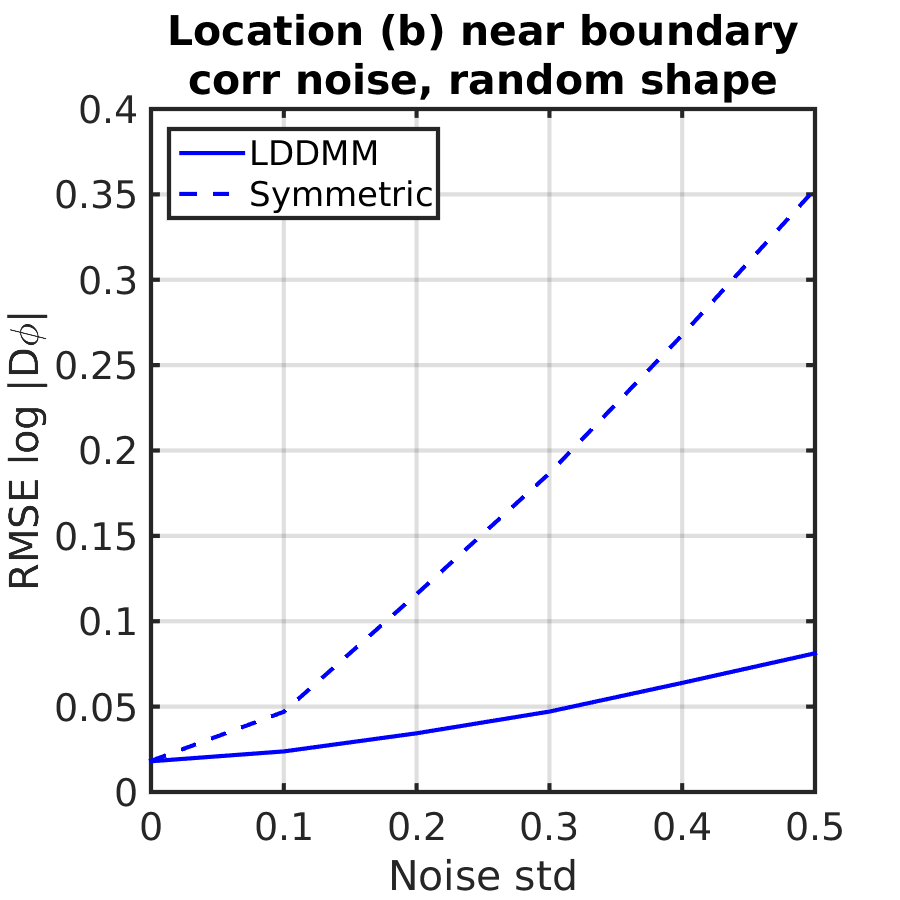}
\includegraphics[width=0.324\textwidth]{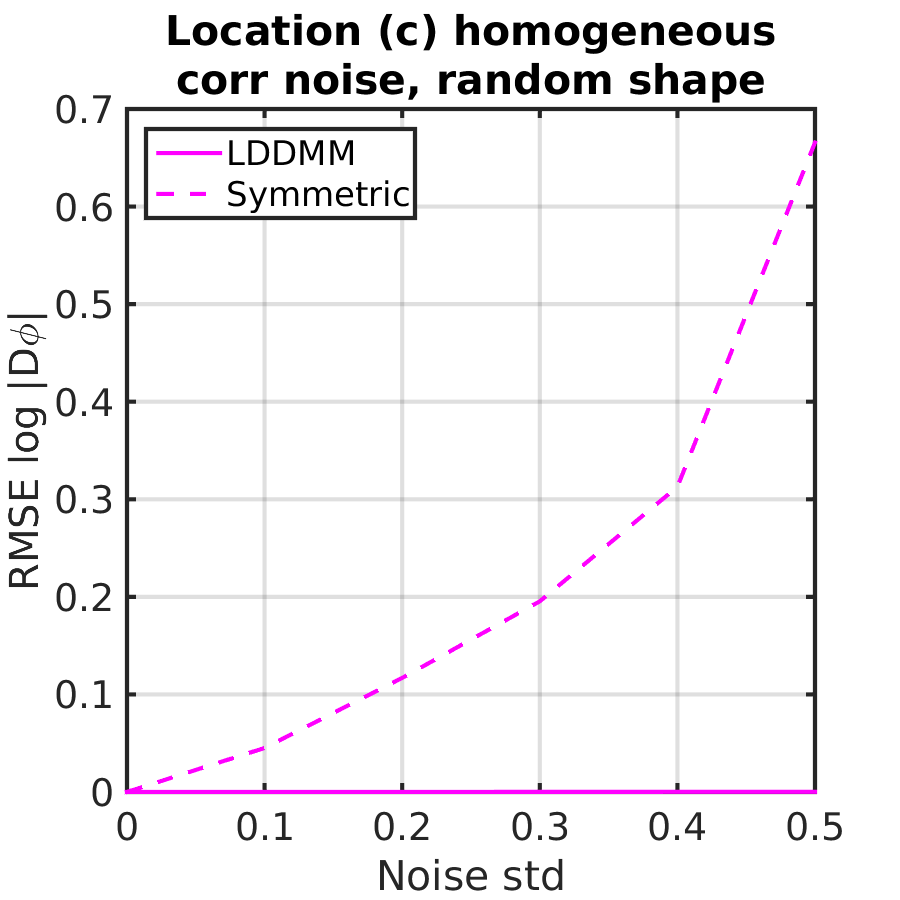}
\caption{\label{fig:curves-corr-identity} RMSE of log canonical volume form in correlated noise;
LDDMM solid, symmetric dashed. Top row shows identity transformation; bottom row shows random transformation.
\label{fig:curves-corr-random} 
}
\end{figure*}

To compare homogeneous regions to boundary regions, we show RMSE in Figs. \ref{fig:curves-white-identity}, \ref{fig:curves-corr-identity} as a function of noise at several regions identified in Fig. \ref{fig:white-noise} (bottom left). 
Results for each of these regions are shown; 
Fig. \ref{fig:curves-white-identity} examines white noise; Fig. \ref{fig:curves-corr-identity} examines correlated noise.

The symmetric method degrades at higher noise levels, as well as with correlated noise.  
In the homogeneous regions the LDDMM has no uncertainty, and as one moves away from gradients the uncertainty quickly drops to zero.
Location (a) illustrates a place of high gradient on the boundary in which we see at reasonable noise levels equivalent performance.
The effect of noise correlation is to increase RMSE, the symmetric method being particularly harshly affected.


\subsection{Cramer-Rao bound}
\begin{figure*}
\centering
\includegraphics[width=0.324\textwidth,clip,trim=1.1in 0in 0.82in 0in]{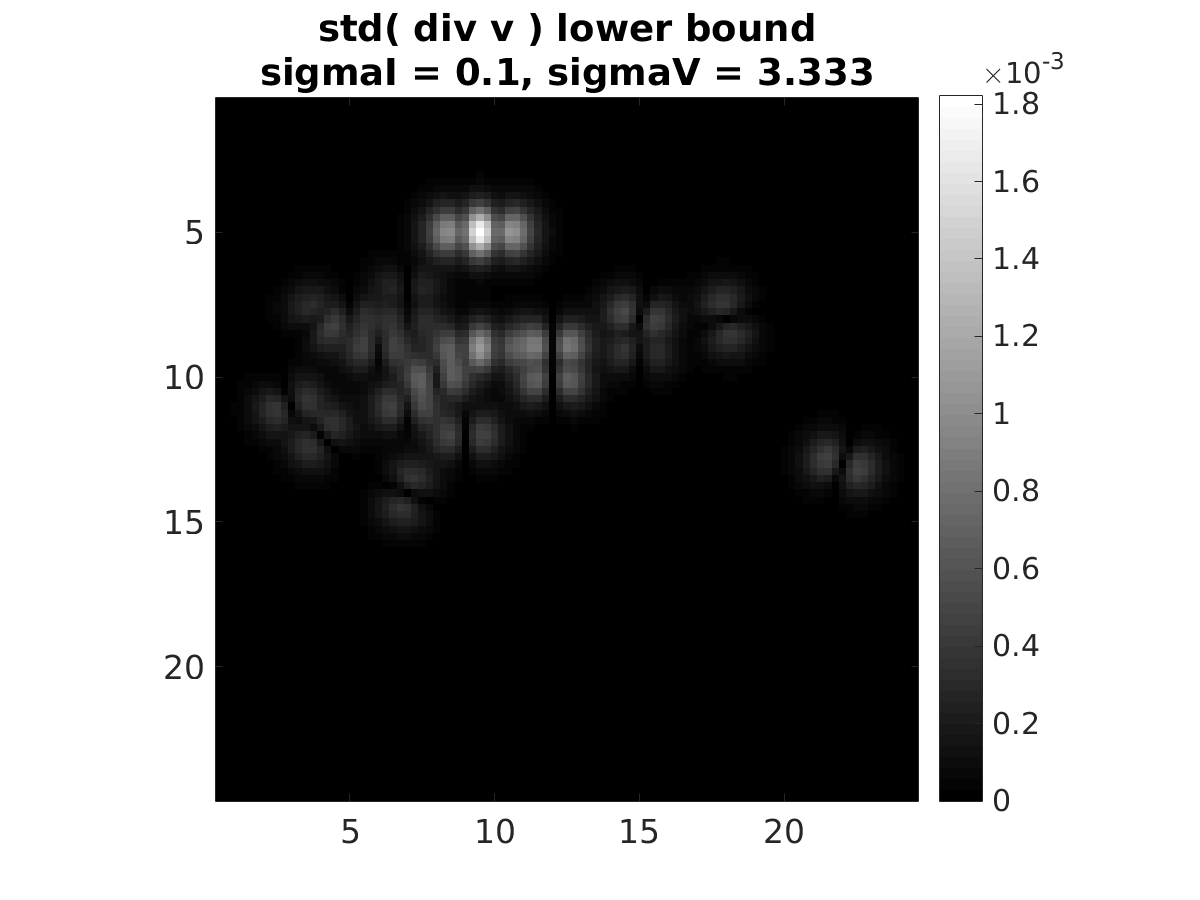}
\includegraphics[width=0.324\textwidth,clip,trim=1.1in 0in 0.82in 0in]{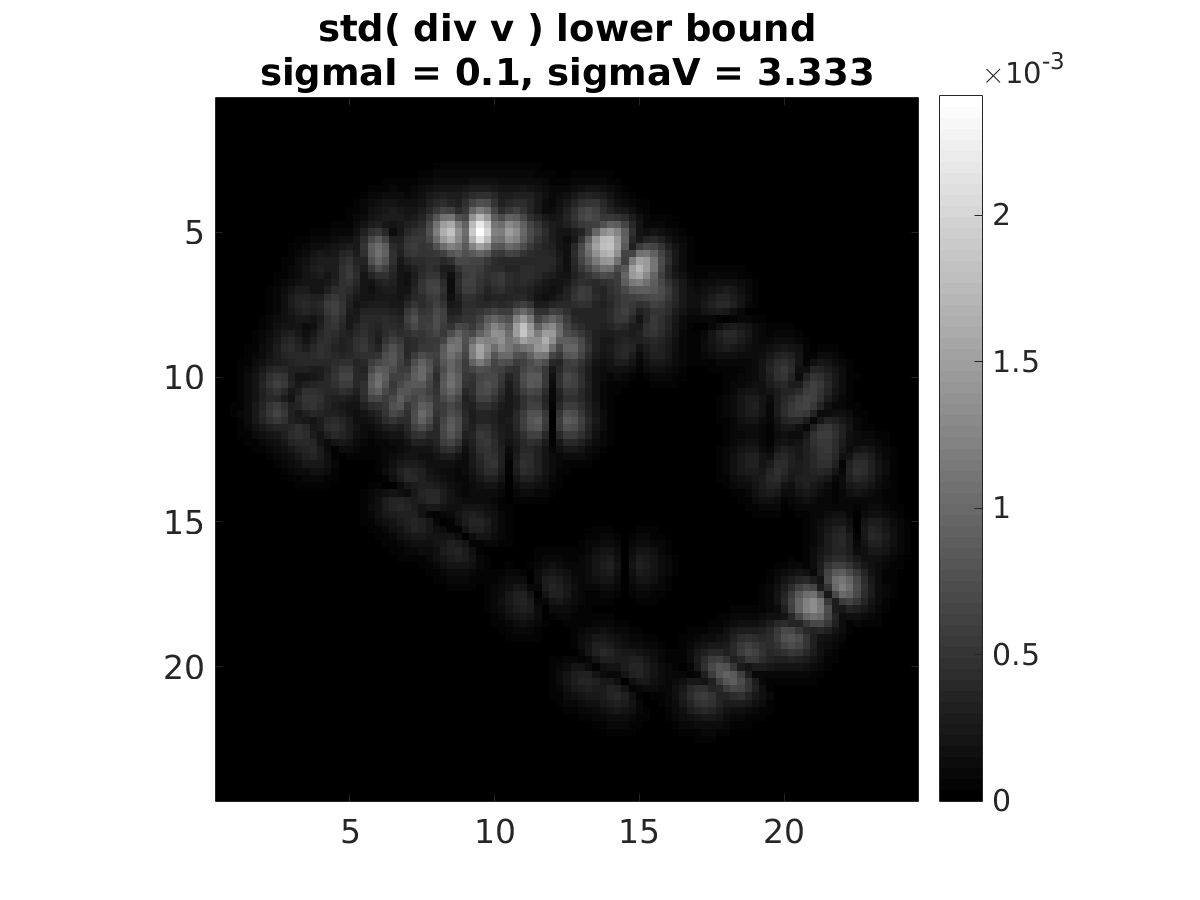}
\includegraphics[width=0.324\textwidth,clip,trim=1.1in 0in 0.82in 0in]{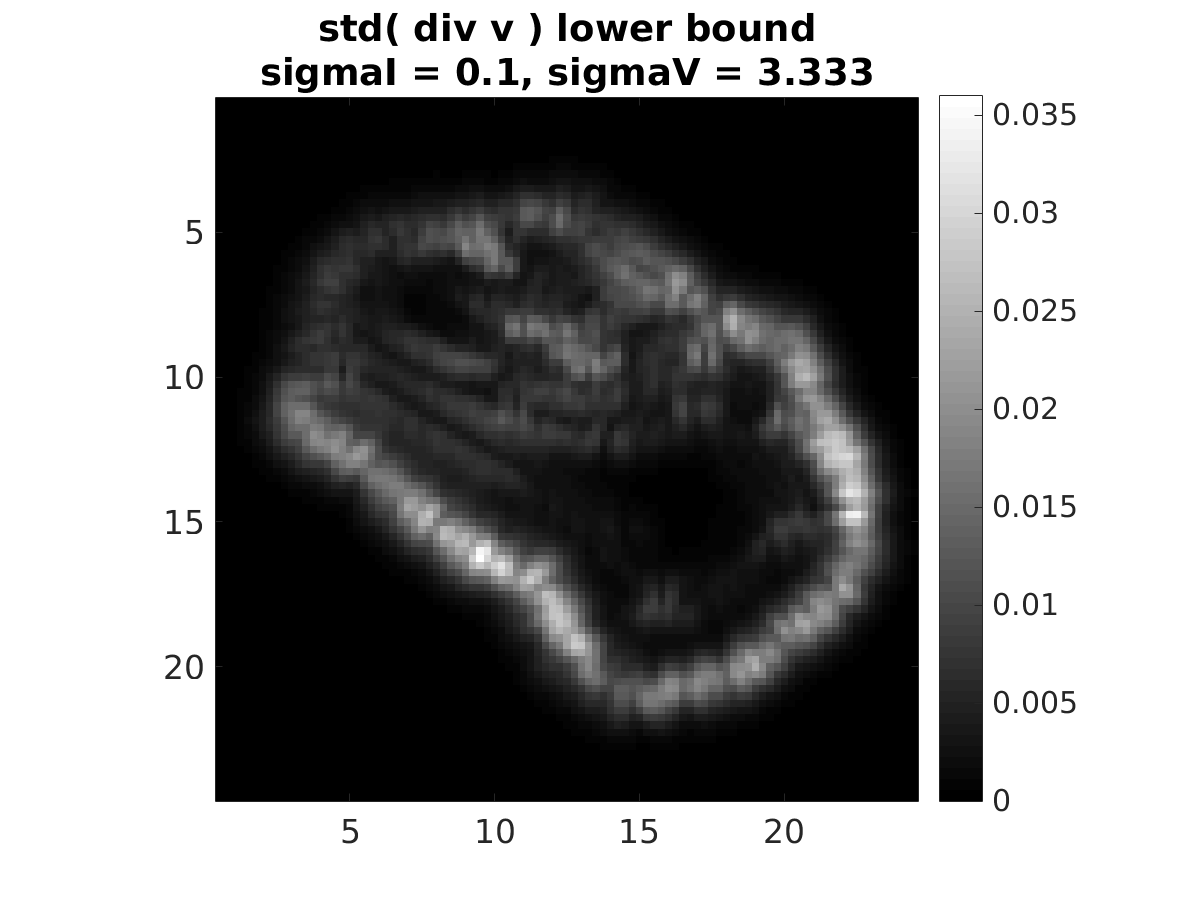}
\\
\includegraphics[width=0.324\textwidth,clip,trim=1.1in 0in 0.82in 0in]{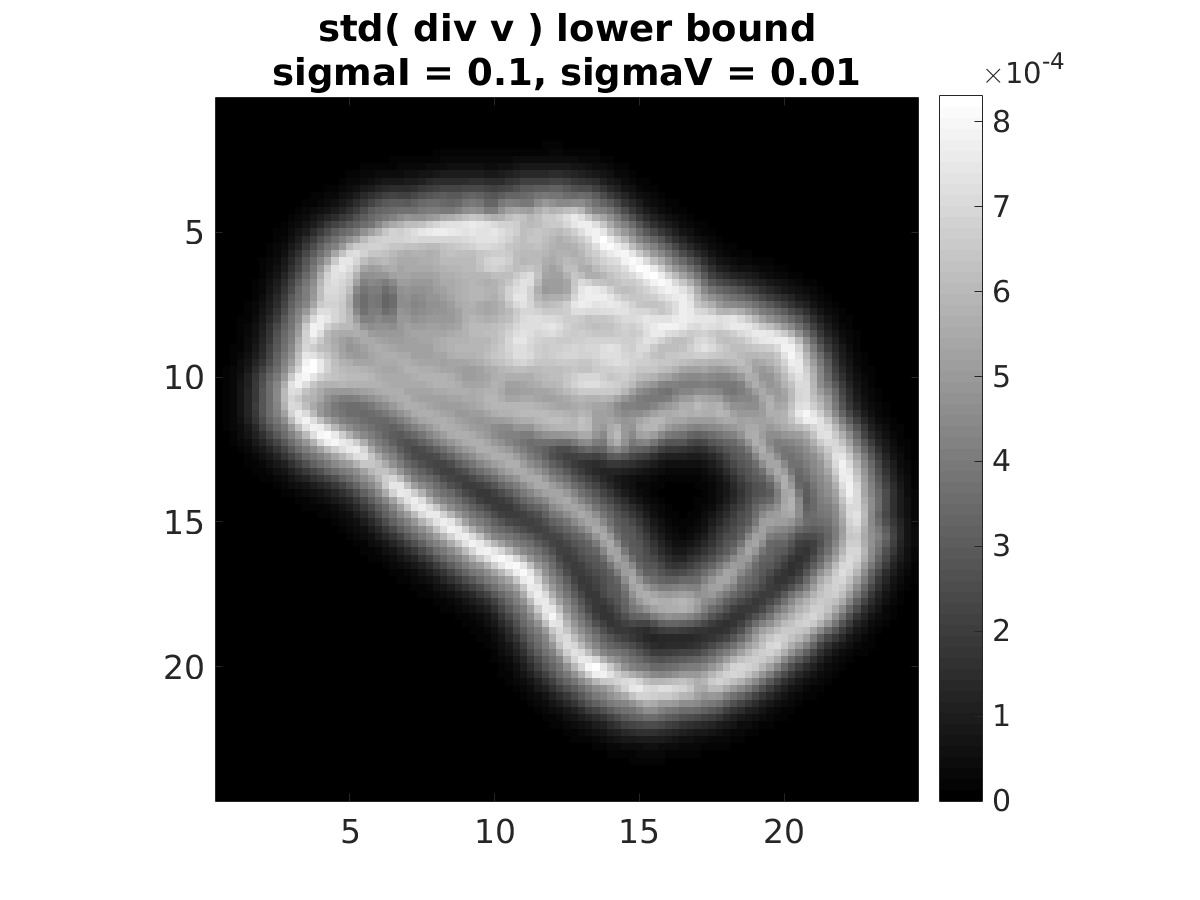}
\includegraphics[width=0.324\textwidth,clip,trim=1.1in 0in 0.82in 0in]{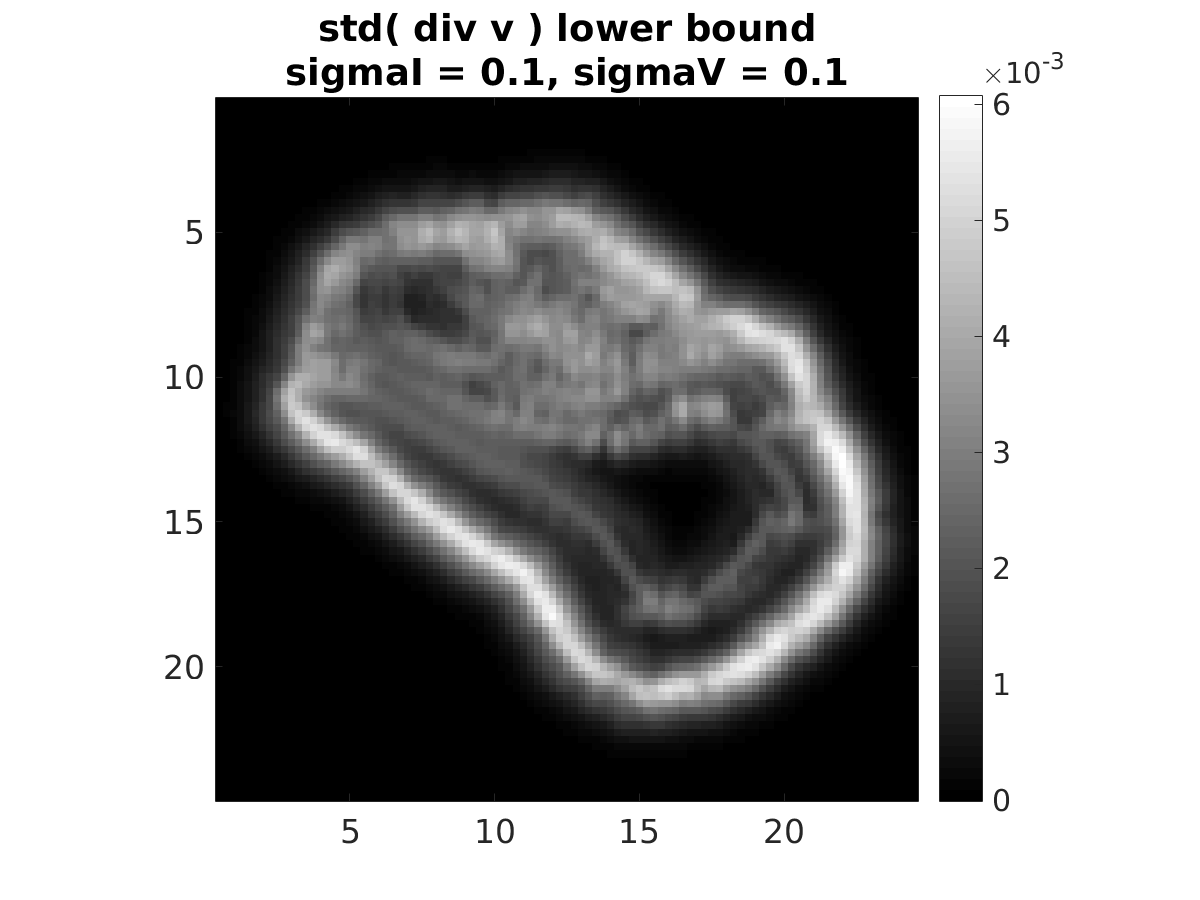}
\includegraphics[width=0.324\textwidth,clip,trim=1.1in 0in 0.82in 0in]{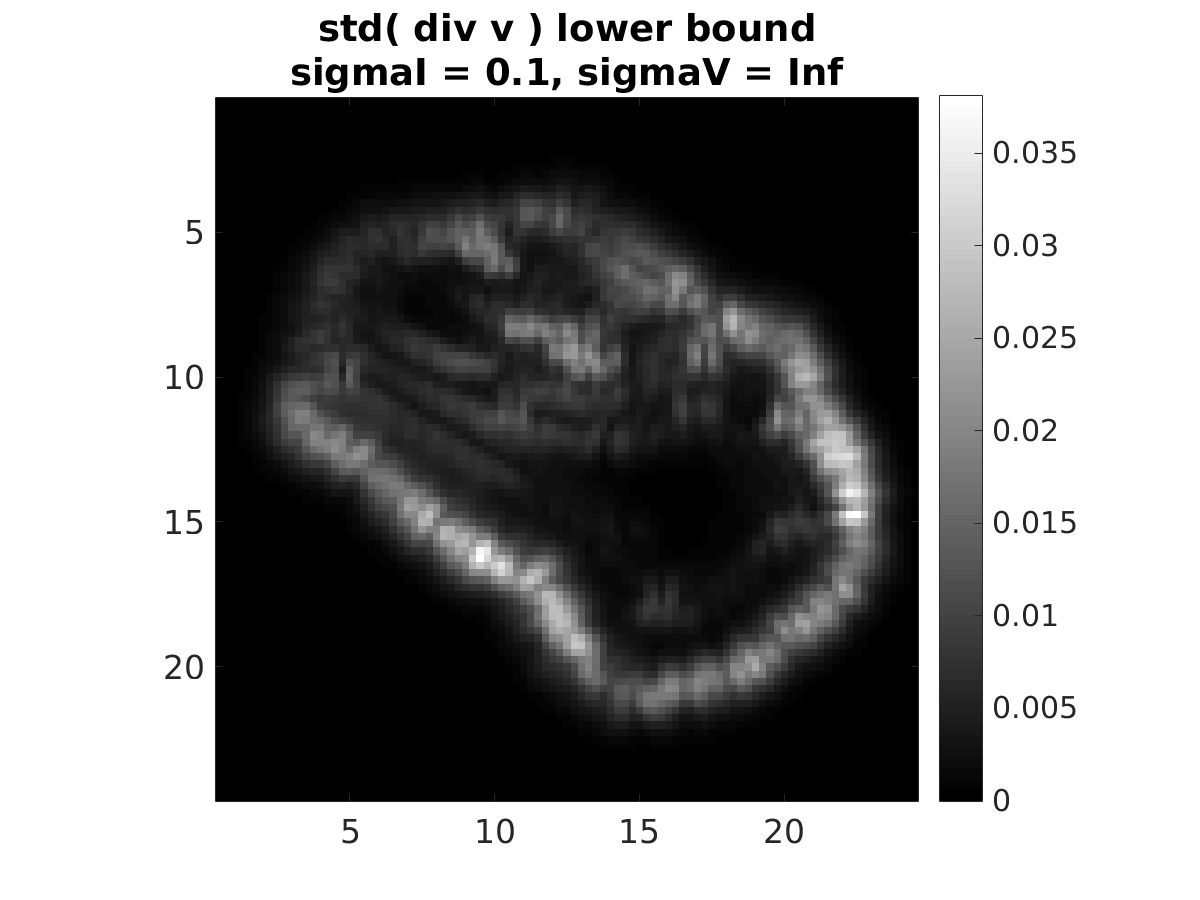}
\caption{\label{fig:CRexpansion} Top row: Finite dimensional expansions for Cramer-Rao bound on $\nabla \cdot \hat v$.  Expansion functions are Green's kernels of $A$ on image boundaries, where the image is 76 expansion functions, 318 expansion functions, 
or 3432 expansion functions.  Bottom row: tradeoff between a strong prior term (left) and a strong data term (right).}

\end{figure*}

Examples of the Cramer-Rao bound on the divergence of $\hat v_n$ for different $n$ are shown in the top row of Fig. \ref{fig:CRexpansion}.  One observers an asymmetry between uncertainty tangent and normal to level lines, which has been noted by other authors \cite{simpson2011longitudinal} and is implied by our consideration of the stabilizer of the diffeomorphism group.  In the bottom row  of Fig. \ref{fig:CRexpansion} we show the lower bound as it varies with changing balance between the image gradient \eqref{CRB-equation} term to the quadratic prior term \eqref{Bayes-bound-equation}.  These results have an important implication for image mapping parameter selection in the presence of noise.  Note that when both terms are multiplied by the same constant, the bound increases linearly and the displayed image will look the same.  The behavior as we transition from ``prior dominant'' to ``image dominant'' can be seen as we move from left to right.

\section{Discussion}

%
%

In this work we demonstrated that image registration algorithms which give comparable accuracy in terms of alignment of observable anatomical boundaries, can lead to widely different estimates of the canonical volume form.  In particular we showed that by employing the non-symmetric procedure associated to geodesic matching of LDDMM, we have explicitly accounted for the stabilizer of the diffeomorphism group, reducing the dimensionality of mappings to that of the observable image boundaries, results in favorable performance in the presence of noise.
Interestingly the small perturbation argument of
Equation \eqref{eq:small-perturbation} also illustrates the importance of the asymmetry, and shows that it is instructive to consider the effect of image variability on estimation of the canonical volume form in 2 scenarios.  
First, because of the null-space of identifiability of motions along level lines, the homogeneous regions that are distant from image gradients by an amount larger than the spatial scale of $K$ have clear identifiability issues in the symmetric methods. This leads to favorable performance of the asymmetric LDDMM method.
Second, boundaries between anatomical structures, such as gray white matter interfaces, have high image gradients.  In these cases the $\delta v$ is approximately related to $\delta J$ through the pseudoinverse of $(\nabla I) (\nabla I)^T$.  

The simulation of noise in the images clearly illustrates that
the departure between source and channel in Shannon's classic model is at the heart of the random orbit model, and implies the importance of asymmetry.
In this setting, the targets are not in the homogeneous orbit of the template under diffeomorphisms. The prior distribution and the space associated to it is separate from the targets as outputs of the noise channel. 
Noisy targets have gradients which are nonzero everywhere, which explains the clear superiority of the asymmetric approach of LDDMM in this situation.

This fills an important gap in our knowledge of how image registration algorithms can or should be used to quantify biological properties of tissue.
%
%
In brain morphomety \cite{ashburner2000voxel}, the canonical volume form at each voxel in a 3D image is used to quantify structural differences between populations, for example measuring atrophy due to disease or aging.  In microscopy it is convenient to idealize cellular locations in terms of spatial point process models \cite{dls91}.  Here the canonical volume used to define a relationship between counts (at the micron scale) and densities (at the mm scale) in a standard atlas coordinate system. During the earliest phases of preclinical Alzheimer's disease for example, when cells have not yet died, cell density changes due to dissolution of the neuropil.
When choosing a mapping procedure, one must consider that algorithms or parameters that may be optimal for registration accuracy, may be considerably sub optimal for quantifying local expansion or contraction.

One challenge here is in considering the effect of the many possible parameter choices that define mapping algorithms. The parameters considered here represent realistic choices for neuroimaging applications, and were chosen based on experience in studying brain morphometry.  An important parameter not explicitly varied was the spatial scale of the transformation's smoothness.  Claims made for ``homogeneous regions'' or ``regions near a boundary'', should be understood as related to this spatial scale.  The MR image considered for our experiment was chosen because it has features at many different spatial scales, from the closely packed structures of the dentate gyrus, to the broadly uniform white matter of the angular bundle.  We have observed experimentally that as the spatial scale of the deformation grows very large, every location in the image becomes ``near a boundary'', and the differences between the two methods considered become less pronounced.





One limitation of this work is that the experiments considered only two dimensional images.  This was for reasons of computational speed when considering statistical ensembles of mapping results, as well as for ease of display and interpretation.  The theoretical developments here do hold in three dimensions, and we expect our findings to generalize.









This work is important for two important biomedical applications.  The first is the effect of data noise on atlas registration procedures.  When considering areas such as surgical planning, the quantification of ``average performance'' may be insufficient.  The consequence of deviations from average in this context can incredibly costly. The second is the extension of these atlas mapping technologies to morphometry, and the study of cell and process densities. 

In the new era of computational anatomy enabled by large volumes of light microscopic data, quantification of the uncertainty in the coordinate mapping between individual data sets and templates is important both for fundamental science applications, and for applications to the study of pathological conditions or computer aided diagnoses.   


\section*{Acknowledgements}
Funding: This work was supported by the National Institutes of Health [P41-EB015909, 
R01-EB020062,
R01-NS102670, 
U19-MH114821, and U01 MH114824 
];  the Kavli Neuroscience Discovery Institute, the Crick-Clay Professorship, CSHL, and the H N Mahabala Chair, IIT Madras.

\appendix


\section{Experimental Parameters}
Parameters used in our simulations are listed in Table \ref{tab:experiment}
\label{experimental-parameter-table}.  


\begin{table*}[!htb]
\caption{\label{tab:experiment} Summary of experimental parameters}
\begin{tabular}{ll}
Parameter & Values\\
\hline \hline
Deformation of atlas & identity, random diffeomorphism\\
Spatial scale of deformation, $a$ & 0.25mm (2 pixels)\\
Approximate magnitude of deformation & 2-3 pixels\\
Image intensity & Binary 0-1 segmentations\\
Noise level & 0, 0.1, 0.2, 0.3, 0.4, 0.5\\
Noise correlation & 0 pixels, 1.5 pixels\\
Matching methods & LDDMM, symmetric LDDMM\\
Matching $\sigma_I$ & 0.1\\
Regularization $\sigma_V$ & 3.33 (0.01, 0.1, $\infty$ shown in Fig. \ref{fig:CRexpansion})\\
Equivalent single parameter $\sigma$ & 0.03\\ 
Matching gradient descent step size & 0.018\\
Number of realizations & 100
\end{tabular}
\end{table*}

\section{Proof of LDDMM Inexact Matching}
\begin{proof}
\label{Euler-Lagrange-appendix}
We will need the perturbation of the inverse. 
Let $\phi^\epsilon = \phi + \epsilon \delta \phi$, computing the variation
uses the fact that $(\phi + \epsilon \delta \phi) \circ (\phi^{-1} + \epsilon \delta \phi^{-1}) = id $ giving
\begin{equation}
\delta \phi^{-1} = - (D \phi)|_{\phi^{-1}}^{-1} \delta \phi |_{\phi^{-1}} \ .
\label{variation-of-inverse-equation}
\end{equation}

Define the total cost $C(\phi)$ for LDDMM to be minimized:
\begin{eqnarray}
\nonumber
C(\phi) = \int_0^1\int_X L(\phi,\dot \phi)dx dt + U(\phi_1) \ , \\
  U(\phi_1) = \frac{1}{2\sigma^2}\int_X |J-I\circ \phi_1^{-1}|^2 dx \ 
\nonumber
\end{eqnarray}
and $L(\phi,\dot \phi)$ is given by
\eqref{Lagrangian-equation}.
The Euler-Lagrange equation follows from first order perturbation, $ \phi \rightarrow 
\phi + \epsilon \delta \phi, \dot \phi \mapsto \dot\phi + \epsilon \frac{d}{dt}\delta \phi  $:
\begin{align*}
&\frac{d}{d \epsilon} C(\phi+\epsilon \delta \phi)|_{\epsilon=0} =
\int_0^1 \int_X \frac{\partial L (\phi,\dot \phi)}{\partial \phi} \cdot \delta \phi dx dt \\
&+ \int_0^1 \int_X \frac{\partial L(\phi,\dot \phi)}{\partial \dot \phi}\cdot \frac{d}{dt}  \delta \phi dx dt \ + \frac{d}{d\epsilon} U(\phi_1+\epsilon \phi_1)|_{\epsilon=0} \ .
\end{align*}
Integrating by parts gives two equations using zero-boundary $\delta \phi_0=0$:
\begin{align}
&\int_0^1 \int_X \left( \frac{\partial L (\phi,\dot \phi)}{\partial \phi} - \frac{d}{dt} \frac{\partial L(\phi,\dot \phi)}{\partial \dot \phi} 
\right) \cdot 
(\delta \phi) dx dt  = 0;
\label{Euler-Lagrange-equation}
\\
&\int_X \frac{\partial L(\phi_1,\dot \phi_1)}{\partial \dot \phi_1} \cdot \delta \phi_1 dx + \frac{d}{d\epsilon} U(\phi_1 + \epsilon \phi_1) |_{\epsilon=0}  = 0\ .
\label{boundary-term}
\end{align}
To calculate the variation of the endpoint term gives
\begin{align*}
&\frac{d}{d\epsilon}|_{\epsilon = 0} U(\phi_1+\epsilon \delta \phi_1) \\
&=  \int_X \frac{1}{\sigma^2} (J - I \circ \phi_1^{-1}) (D \phi_1)_{\phi_1^{-1}}^{-1T} (\nabla I)_{\phi_1^{-1}} \cdot \delta \phi_1 |_{\phi_1^{-1}} dx
\nonumber
\\
&=
\int_X\frac{1}{\sigma^2} (J \circ \phi_1 - I )  (D \phi_1)^{-1T} \nabla I  |D \phi_1 | \cdot \delta \phi_1  dx
\ .
\nonumber
\end{align*}
Using the fact that the conjugate momentum is $p= \frac{\partial L}{\partial \dot \phi}= [Av] \circ \phi |D\phi|$, then
Equation \eqref{Euler-Lagrange-equation}
is the Euler-Lagrange equation;
$$
\frac{d}{dt} p_t   +(Dv_t)^T\circ \phi_t \ p_t =0 \ ; $$
\eqref{boundary-term} is the boundary matching term
reducing to
\begin{eqnarray}
&&
 \int (p_1 +\frac{1}{\sigma^2} (J \circ \phi_1 - I )  (D \phi_1)^{-1T} \nabla I )  |D \phi_1 |\cdot \delta \phi_1  dx = 0
\ .
\nonumber
\end{eqnarray}
giving
\begin{equation}
p_1 +\frac{1}{\sigma^2} (J \circ \phi_1 - I )  (D \phi_1)^{-1T} \nabla I   |D \phi_1| = 0
\ .
\end{equation}

The result for $p_t$ in terms of $p_1$ is obtained by applying the result of Appendix \ref{app:EL} twice. First calculate at time 0: $p_0 = D\phi_1^T p_1$, then at time $t$: $ p_t = D\phi_t^{-T}D\phi_1^T p_1 $, giving the result
\begin{equation}
p_t + \frac{1}{\sigma^2}(J\circ\phi_1 - I)D\phi_t^{-T}\nabla I |D\phi_1| = 0
\end{equation}
\end{proof}

\section{Euler-Lagrange and Conservation are equivalent}
\begin{proof}
\label{app:EL}
\begin{eqnarray}
&&
 \dot p_t   +(Dv_t)^T\circ \phi_t \ p_t  =0 \ ; 
 \label{Euler-lagrange-appendix}
\\
&& D\phi_t^T p_t  =   p_0 \ .
\label{conservation-conjugate-momentum}
\end{eqnarray}

To see this,
take the derivative
\begin{align*}
\frac{d}{dt} D \phi_t^T p_t &=
D(v_t \circ \phi_t)^T p_t + D \phi_t^T \frac{d}{dt} p_t \notag\\
&= D \phi_t^T Dv_t^T \circ \phi_t p_t +D \phi_t^T \dot p_t \ .
\end{align*}
The last equality is zero by Euler-Lagrange equation \eqref{Euler-lagrange-appendix}.


\end{proof}

\section{Small deformation momentum}
\label{Proof-of-small-deformation-momentum}
\begin{proof}
Let $\phi^\epsilon = \phi + \epsilon \delta \phi (\phi)$, then $v^\epsilon = v+ \epsilon \delta \phi (\phi) $.
Computing the variation requires perturbation of the inverse, $\delta \phi^{-1} = - (D \phi)|_{\phi^{-1}}^{-1} \delta \phi |_{\phi^{-1}} \ $ determined in \eqref{variation-of-inverse-equation}.

Then we have
\begin{align}
&\frac{d}{d \epsilon} C(\phi+\epsilon \delta \phi) |_{\epsilon = 0} \nonumber\\
&=  \frac{d}{d\epsilon} \frac{1}{2 } \int_X A v^\epsilon \cdot v^\epsilon dx +\frac{1}{2 \sigma^2} \| J- I \circ \phi^{\epsilon -1} \|^2 |_{\epsilon=0}
\nonumber\\
&=
\int_X Av \cdot \delta \phi (\phi) \, dx \nonumber\\
&\qquad +  \int_X \frac{1}{\sigma^2} (J - I \circ \phi^{-1} )  D (\phi^{-1})^T ( \nabla I)_{\phi^{-1}} \cdot \delta \phi  \, dx
\ .
\end{align}
Substituting into the second integral $x=\phi(y)$ with $dx =|d\phi| dy$ gives
us \eqref{momentum-small-deformation}.
\end{proof}

\section{Effect of small perturbations}
\label{proof-of-small-perturbation}
\begin{proof}
We consider the perturbation $J \mapsto J + \epsilon \delta J$ and $v \mapsto v + \epsilon \delta v$ on  \eqref{momentum-small-deformation}, seeking a relationship between the two to first order in $\epsilon \in \mathbb{R}$.
\begin{align*}
&\frac{d}{d\epsilon}[Av + \epsilon A\delta v] \bigg|_{\epsilon = 0} = \frac{d}{d\epsilon}\frac{1}{\sigma^2}(I - (J + \epsilon \delta J)\circ(\phi + \epsilon \delta v)) \\
&\qquad D[\phi + \epsilon \delta v]^{-T}\nabla I |D\phi + \epsilon \delta dv\bigg|_{\epsilon = 0}
\end{align*}
The left hand side is simply $A\delta v$.  The right hand side includes a product of three terms.  The variation of the first gives $- \delta J(\phi) - DJ(\phi)\delta v$.  That of the second gives $-[D\phi]^{-T}D\delta v^T [D \phi]^{-T}$, while the third gives $\text{div}[\delta v]|D\phi|$.  Combining these gives
\begin{align*}
A\delta v &= -\frac{1}{\sigma^2}[DJ(\phi)\delta v + \delta J(\phi)][D\phi]^{-T}\nabla I |d\phi|\\
&- \frac{1}{\sigma^2}[I - J(\phi)][D\phi]^{-T}D\delta v^T [d\phi]^{-T} \nabla I |d\phi|\\
& +\frac{1}{\sigma^2}[I - J(\phi)][D\phi]^{-T}|D\phi|\text{div}[\phi]
\end{align*}

We consider one of the cases examined experimentally, with $I = J$ and therefore $\phi = id$ (identity).  Only the first term is nonzero, giving
\begin{align*}
A\delta v = \frac{1}{\sigma^2}[DI \delta v + \delta J]\nabla I 
\end{align*}
Rearranging gives the relationship \eqref{eq:small-perturbation}.
\end{proof}

\bibliography{LDDMM}

\end{document}